\documentclass{article}[11pt,a4paper]

\usepackage{color}      
\usepackage{amsmath,amssymb,amsfonts,bbm,bm,graphicx,hyperref,times,subfigure}
\usepackage[T1]{fontenc}
\usepackage{dsfont}

\newcommand{\be}{\begin{equation}}
\newcommand{\ee}{\end{equation}}
\newcommand{\bea}{\begin{align}}
\newcommand{\eaa}{\end{align}}
\newcommand{\id}{\mathbbm{1}}

\newcommand{\ket}[1]{|{#1}\rangle}
\newcommand{\bra}[1]{\langle{#1}|}
\newcommand{\sig}{{\boldsymbol{\sigma}}}
\newcommand{\tr}[1]{{\rm Tr}\left[{#1}\right]}

\usepackage{epstopdf}
\usepackage{subfigure}

\title{\bf \textsf{Conditional and unconditional Gaussian quantum dynamics}}
\author{
Marco G. Genoni,${}^{a}$ 
Ludovico Lami,${}^{b}$ and Alessio Serafini${}^{a}$\\
\small${}^{a}$Department of Physics \& Astronomy, University College London, \\
\small Gower Street, London WC1E 6BT, UK;\\
\small${}^{b}$ F\'{i}sica Te\`{o}rica: Informaci\'{o} i Fen\`{o}mens Qu\`{a}ntics, Departament de F\'{i}sica, \\ 
\small Universitat Aut\`{o}noma de Barcelona, 08193 Bellaterra (Barcelona), Spain.}

\begin{document}

\maketitle

\begin{abstract}
This article focuses on the general theory of open quantum systems in the Gaussian regime and explores a number of diverse ramifications 
and consequences of the theory. We shall first introduce the Gaussian framework in its full generality, including a classification 
of Gaussian (also known as ``general-dyne'') quantum measurements. In doing so, we will give a compact proof for the parametrisation of the most general 
Gaussian completely positive map, which we believe to be missing in the existing literature. We will then move on to consider the 
linear coupling with a white noise bath, and derive the diffusion equations that describe the evolution of Gaussian states under such circumstances.
Starting from these equations, we outline a constructive method to derive general master equations that apply outside the Gaussian regime. 
Next, we include the general-dyne monitoring of the environmental degrees of freedom and recover 
the Riccati equation for the conditional evolution of Gaussian states. Our derivation relies exclusively on 
the standard quantum mechanical update of the system state, through the evaluation of Gaussian overlaps. 
The parametrisation of the conditional dynamics we obtain is novel and, at variance 
with existing alternatives, directly ties in to physical detection schemes. We conclude our study with two examples of 
conditional dynamics that can be dealt with conveniently through our formalism, demonstrating how monitoring 
can suppress the noise in optical parametric processes as well as stabilise systems subject to diffusive scattering.
\end{abstract}

\section{\bf \textsf{Motivation, background, and plan of the paper}}

As any boater knows, the continuous observation and steering of a physical system is an obvious way to achieve its 
dynamical control. 
Slightly more subtly, the mere act of observing and gaining information 
also typically reduces the entropy content of a system.
Both dynamical control and entropy reduction -- which is essentially what `cooling' protocols aim at -- are primary objectives towards the realisation of more and more advanced experiments in the quantum regime and, ultimately, quantum technologies, 
and the expedient description of continuously observed (``monitored'') quantum systems is hence currently of great interest. 

However, the exact treatment of the conditional dynamics of quantum systems ({\em i.e.}, of dynamics where the evolution of the system is 
conditioned by the occurrence of certain measurement outcomes) is typically rather involved, 
as it requires one to incorporate the irreversible update prescribed by the Born rule and by the projection postulate into the evolution.
Several approaches are available to this aim, such as the stochastic Schr\"odinger and 
master equations \cite{carmichael,gardiner,WisemanMilburn}, quantum stochastic calculus \cite{belavkin92} 
or the quantum jumps formalism \cite{plenioknight}.
In dealing with weak continuous measurements, 
{\em i.e.}~with quantum measurements which imply little disturbance on the system, 
being realisable by coupling it to a probe for an infinitesimal time interval 
and then by measuring the latter continuously in time, the treatment through stochastic Schr\"odinger equations is 
well suited. Such a treatment, also known as the quantum trajectories approach, 
leads to a modified Schr\"odinger equation where stochastic increments take into account the 
probabilistic effect of the measurement on the evolving state. Through the definition of stochastic master equations, this 
treatment can also be extended to evolving mixed states.
Nonetheless, the analytic integration of such equations is typically impossible, and their derivation may prove difficult to some potential users, 
as it requires a certain familiarity with stochastic calculus \cite{gardiner}.

There is one notable exception to this state of affairs: the case of Gaussian diffusive dynamics, which covers the broad range 
of situations where the following conditions are met: 
\begin{itemize}
\item an open quantum system couples linearly to its environment;
\item system and environment are governed by Hamiltonians which are at most quadratic in their canonical
operators;
\item the environment is continuously monitored through Gaussian measurements, in a sense to be 
specified in the following;
\item the initial state is a Gaussian state. 
\end{itemize}
These conditions may seem rather restrictive, but are actually ordinarily met by a vast number of 
existing experimental set-ups in the areas of quantum optics, trapped ions, opto-mechanics, atomic ensembles and 
certain superconducting degrees of freedom, whenever finite dimensional degrees of freedom or anharmonicities 
are not involved.\footnote{Note that this also excludes the spin-boson model and its variations.} 
What we shall refer to as Gaussian measurements are also customarily carried out 
in laboratories with comparatively high efficiency, while the description of quantum noise, as we will see, is a natural part of the Gaussian picture.

What is perhaps even more important to remark is that the restriction to the Gaussian realm, while obviously not capable to capture the 
full wealth of dynamics allowed in the Hilbert space, still allows one to include most of the processes relevant to quantum technologies, 
such as squeezing (whereby certain canonical quadratures have uncertainties below the vacuum state noise, and can hence 
be used in precision measurements), quantum entanglement (stronger than classical non local correlations,
abundant in Gaussian two-mode squeezed states), 
and cooling (where the entropy is drained out of a system in order to reach a pure quantum state, 
and initialise information protocols or low noise experiments). The Gaussian restriction sketched above, which 
over the years has carved for itself a dedicated 
niche within the research on quantum information \cite{jensmartin,gerryfaber,gaussianbucco,rmp}, has often been criticised on the grounds that no real 
genuinely quantum effects can ever be observed without leaving it as some point. This is indeed the case, since 
Gaussian states can be mimicked by classical probability distributions.\footnote{More specifically, here we are referring to the fact that the 
measurement statistics resulting from {\em Gaussian measurements} on Gaussian states can always be reproduced with classical systems, 
and thus do not allow for stronger-than-classical correlations.} 
However, the object described by a Gaussian state 
does entail a genuinely quantum description in a Hilbert space. Hence, for instance, 
the preparation of a pure Gaussian quantum state is the preparation of a pure vector of the Hilbert space. 
Likewise, a highly squeezed Gaussian state prepared for a metrological protocol is a state with a very low noise in a certain physical observable
that does grant sub shot-noise precision \cite{caves,sirgeno}, 
regardless of whether it can be mimicked by a classical distribution or not.
The testing and exploitation of Gaussian quantum non-locality does instead require a departure from Gaussian 
measurements although, since certain non-Gaussian measurements are customarily implemented with current technology, 
it is still relevant to study in detail the creation of such a resource.

When the Gaussianity conditions listed above are met, the analysis of monitored, conditional quantum dynamics 
simplifies substantially, and exact analytical formulae are available. 
This fact is very well known, and the instances where these solvable conditional dynamics have been applied in the 
quantum control and quantum optics literature is beyond count \cite{WisemanMilburn}. 
However, notwithstanding the wealth of studies 
they underpinned, the coverage of such Gaussian dynamics found in the literature,
while extensive and general, is always aimed at specialist 
researchers, and typically assumes a certain acquaintaince with stochastic calculus 
and advanced familiarity with the language and notions of either mathematical physics or quantum optics 
(depending on which strand of literature one is tackling). 
We believe a simpler rendition of this specific subject to be possible, and we shall try our hand at it in this article. 
To make our treatment self-sufficient and more readable, the discussion of conditional Gaussian dynamics will be 
embedded into a general treatment of Gaussian dynamical maps.
Let us note that, as a further pedagogical byproduct of our approach, we will sketch a novel general method to derive 
master equations that apply even when the system Hamiltonian is arbitrary 
and not quadratic, and the initial state is not Gaussian and which is, we argue, more straightforward than 
existing derivations.

Note also that, although some of our results apply beyond the class of Gaussian states, 
this is not, as our title clearly indicates, an article on general continuously monitored dynamics,
but rather on their much more specific Gaussian restriction. An excellent, broader introduction to the topic of continuous monitoring has already appeared on this journal, and we gladly refer the reader to it \cite{JacobsCP}.

The plan of the article is as follows: in Sec. \ref{s:Gaussian} we review the properties of Gaussian states, Gaussian unitary evolutions and Gaussian (general-dyne) measurements. In Sec. \ref{s:GaussianCPmap} we review and re-derive the main properties of Gaussian completely positive (CP) maps, specifying the role of the dual maps in the description of noisy measurements. In Sec. \ref{s:openGaussian} we show how to obtain the evolution equations describing open Gaussian dynamics by considering the interaction with a large Markovian (memoryless) bath. In Sec. \ref{s:filtering} 
we derive the dynamics corresponding to diffusive quantum filtering, that is obtained by continuously monitoring the environment via general-dyne detection. We will show how our formalism can be easily used, by providing 
the reader with two case studies in Sec. \ref{s:examples}. 
We conclude the paper in Sec. \ref{s:conclusions} with a summary and some general remarks.

\section{\bf \textsf{Gaussian quantum states: basic notions and description}} \label{s:Gaussian}
In this article, we will be concerned with continuous variable quantum systems, {\em i.e.}~with quantum degrees of freedom that encompass observables with real continuous spectra. Although such systems are prominent in the traditional pedagogy 
of quantum mechanics, since the motional degrees of freedom of particles are described by pairs of such observables, 
and should hence be familiar to the vast majority of readers with a background in physics, 
they have been typically the underdogs within the quantum information literature, given its emphasis on mimicking classical 
digital systems, a task that only requires finite dimensional Hilbert spaces.

A ``continuous'' variable quantum system, as opposed to a ``discrete'' one endowed with a finite dimensional Hilbert space,
is usually defined by introducing pairs of self-adjoint canonical operators 
$\hat{x}_j$ and $\hat{p}_j$, for $j= 1,\dots,n$, which, if recast as a vector 
${\bf \hat{r}} = (\hat{x}_1,\hat{p}_1,\dots,\hat{x}_n,\hat{p}_n )^{\sf T}$, satisfy the canonical commutation relations:
\be \label{ccr}
[\hat{r}_j,\hat{r}_k] = i \Omega_{jk} \:,
\ee
where the matrix
\be
\Omega = \bigoplus_{j=1}^n \omega \:, \:\:\: \textrm{with} \:\: \omega = \left(
\begin{array}{ c c}
0 & 1 \\
-1 & 0
\end{array}
\right), 
\ee
is referred to as the symplectic form, for reasons that will become apparent in the following.
At times, we will adopt the handy convention whereby we shall not specify the dimension (number of modes) of $\Omega$:
the symbol $\Omega$ without a label will always stand for the anti-symmetric form of the dimension given by the matrix it multiplies to the left and/or right. 
When we will deem it expedient to clarity, we shall instead explicitly specify the number of number of degrees of freedom $k$ through 
labelling, as in $\Omega_k$. 
This arrangement will be useful when dealing with composite systems, comprising subsystems with possibly different number of modes. 
The couples $\{\hat{x}_j,\hat{p}_j\}$ might refer to the positions and momenta of material particles in first quantization,
or to the field operators of a bosonic field in second quantization, such as the magnetic and electric quadratures of the 
electromagnetic field.

In the following we will deal with Hamiltonians at most quadratic in the canonical operators, {\em i.e.} 
that can be written as
\be
\hat{H} = \frac12 \hat{\bf r}^{\sf T} H \hat{\bf r} + \hat{\bf r}^{\sf T} {\bf r}_{H} \:, \label{eq:quadratic}
\ee
where ${\bf r}_H$ is a $2n$-dimensional real vector and $H$ a symmetric matrix, known as the Hamiltonian matrix.

Quadratic Hamiltonians are intimately related to the subset of ``Gaussian'' states, which are in a certain sense 
the quantum analogue of multivariate Gaussian distributions in classical probability theory. 
The set of Gaussian states may be defined as the set of all the ground and thermal states of (at most) quadratic Hamiltonians with positive definite Hamiltonian matrix, {\em i.e.} a state $\varrho_G$ is Gaussian if and only if 
there exist a symmetric, real, positive definite $H$, a ${\bf r}_{H} \in {\mathbbm R}^{2n}$ and a $\beta\in {\mathbbm R}^{+}$, 
such that 
\be
\varrho_G = \frac{e^{-\beta \hat{H}}}{\tr{e^{-\beta \hat{H}}}}\:, \label{gsgs}
\ee
with $\hat{H}$ defined as in Eq.~(\ref{eq:quadratic}).
Notice that this definition of a Gaussian state is entirely equivalent to the standard one found in the 
quantum optics and quantum information literature, where a Gaussian state is defined 
as a state with a Gaussian characteristic or Wigner function \cite{gaussianbucco}. Note also that the definition includes the 
limit $\beta\rightarrow +\infty $, where one recovers the pure ground state of the quadratic Hamiltonian, 
which is a Gaussian state too.

The (`symmetrically ordered') characteristic function of any quantum state $\varrho$ is defined as \cite{barnettradmore}
\be
\chi({\bf r}) = \tr{\hat{D}_{-{\bf r}} \varrho}
\ee
where
\be
\hat{D}_{\bf r} = e^{i {\bf r}^{\sf T} \Omega \hat{\bf r} } \label{eq:weyl}
\ee
is the so-called Weyl (displacement) operator. 
By virtue of the following Fourier-Weyl relation \cite{Glauber} which, in a loose sense,  
states that displacement operators form an orthogonal basis in the space of bounded operators,
the density operator $\varrho$ describing the state of the $n$-mode continuous variable system can be written as 
\be
\varrho = \frac{1}{(2\pi)^n} \int_{\mathbbm{R}^{2n}} {\rm d}^{2n}{\bf r} \:\chi({\bf r}) \hat{D}_{\bf r} \:, \label{multifw}
\ee
where ${\rm d}^{2n}{\bf r} = {\rm d}x_1 {\rm d}p_1 \dots {\rm d}x_n {\rm d}p_n$.
Note that the displacement operators are orthogonal with respect to the Hilbert-Schmidt inner product, in the sense that 
\be\label{weylortho}
{\rm Tr}\left[\hat{D}_{{\bf r}}\hat{D}^{\dag}_{{\bf s}}\right] = (2\pi)^n \delta^{2n}({\bf r}-{\bf s}). 
\ee
Since $\hat{D}_{\bf 0}=\id$, 
this formula may be used to show that $\chi({\bf 0})$ must equal $1$ for the Fourier-Weyl relationship (\ref{multifw}) 
to be consistent with ${\rm Tr}\left[\varrho\right]=1$. 
Notice also that $\hat{D}^{\dag}_{{\bf r}}=\hat{D}_{-{\bf r}}$.

As anticipated above, the characteristic function of a Gaussian state $\varrho_G$ can be written as 
a multivariate Gaussian of $2n$ variables:
\be
\chi_G({\bf r}) = e^{-\frac14 {\bf r}^{\sf T} \Omega^{\sf T} \sig \Omega {\bf r} } e^{i {\bf r}^{\sf T} \Omega^{\sf T} {\bf r}^\prime} \:,  \label{chara}
\ee
where 
\begin{align}
{\bf r}^\prime &= \tr{\varrho_G \hat{\bf r}}  \:, \\
\sig &=  \tr{ \left\{ ( \hat{\bf r} - {\bf r}^\prime ), (\hat{\bf r} -{\bf r}^\prime )^{\sf T} \right \} \varrho_G} \:, \label{CM}
\end{align}
are respectively the vector of first moments and the covariance matrix (CM). In Equations like (\ref{CM}), the quantity inside the anti-commutator has to be 
taken as an outer product: for instance, in components, one would have $\sigma_{jk}=\tr{ \left\{ ( \hat{r}_j - r^\prime_j ), (\hat{ r}_k -{r}^\prime_k )^{\sf T} \right \} \varrho_G}$.
These two set of real quantities univocally describe the Gaussian state $\varrho_G$. 
While the vector of first moments is any unconstrained real vector, 
let us remark that a real symmetric matrix $\sig$ is the CM associated to a quantum state (and, in particular, of a Gaussian state), 
if and only if it satisfies the Robertson-Schr\"odinger uncertainty relation, in the form of the inequality \cite{Simon}
\be
\sig + i \Omega \geq 0 \:. \label{eq:uncertainty}
\ee
In the phase space picture, inspired by classical Hamiltonian dynamics and rigorously defined by quantum quasi-probability distributions 
-- essentially, the Fourier transforms of characteristic functions -- the first moments determine the centres of the Gaussian distributions 
corresponding to Gaussian states, while the covariance matrices describe their shapes.

Because of the very definition of a Gaussian state, if one considers an $(m+n)$-mode case, 
the reduced state of the subsystem described by, say, the first $m$ modes which, in the full quantum mechanical description,
is obtained by taking the partial trace over the last $n$ degrees of freedom, is still a Gaussian state.
The first moments of such a state are simply given by the relevant entries in the full vector of first moments 
(the first $2m$ entries, in this instance), while its covariance matrix is just given by the principal submatrix 
describing the modes of interest (containing the first $2m\times 2m$ entries, in this case). 
This ease in the evaluation of the partial trace, which comes down to selecting specific subvectors and principal submatrices, 
is a major advantage of the Gaussian description, which we shall exploit in the following. 
This simplicity stems from the fact that 
tensor products translate into direct sums in the phase space picture, and from the expediency in evaluating 
marginal distributions of multivariate Gaussian distributions.

Before moving on, let us also remind the reader the integration rule of a multivariate Gaussian, which will be useful 
in the following.  Given a symmetric, real, positive definite matrix $2n\times 2n$ $A$, 
and a $2n$-dimensional vector ${\bf b}$, one has:
\be\label{gaunt}
\int_{{\mathbbm R}^{2n}} {\rm d}{\bf r}\, {\rm e}^{-{\bf r}^{\sf T} A {\bf r}+{\bf r}^{\sf T}{\bf b}} = \frac{\pi^{{n}}}{\sqrt{{\rm Det}\,A}}\, {\rm e}^{\frac14 
{\bf b}^{\sf T}A^{-1}{\bf b}} \; ,
\ee
where the shorthand notation ${\rm d}{\bf r}$ indicates the product of differential of the $n$ integration variables that compose the vector ${\bf r}$.
The equation (\ref{gaunt}) may be applied, along with (\ref{multifw}), (\ref{weylortho}) and (\ref{chara}), to obtain \cite{purity}
\be
{\rm Tr}\left[\varrho^2\right] = \frac{1}{\sqrt{{\rm Det}\sig}} \; . \label{puri}
\ee
The quantity ${\rm Tr}\left[\varrho^2\right]$ reaches its maximum, $1$, for pure states, that is for density matrices that can be written 
as a projector on a vector of the Hilbert space: $\varrho=\ket{\psi}\bra{\psi}$. It is hence an expedient way, related to 
linearised notion of entropy, the Renyi-2 entropy \cite{renyi}, to characterise the purity of a quantum state,
and is especially easy to evaluate for Gaussian states, where it depends only on the determinant of the CM. A Gaussian state is 
pure if and only if the determinant of its CM is $1$. An example of a pure Gaussian state is the vacuum state, as the ground state of the 
free quadratic Hamiltonian is usually referred to: the CM of the vacuum is the identity in our convention. Note that all pure Gaussian states, 
with ${\rm Det}\sig=1$, are ground states of a quadratic Hamiltonian, {\em i.e.} they correspond to the case $\lim_{\beta\rightarrow +\infty}$ in 
Eq.~(\ref{gsgs}). 

\subsection{\bf \textsf{Gaussian unitary dynamics}}
Because of the definition of the set of Gaussian states, the most general unitary dynamics that preserves the Gaussian character 
of a state is generated by a (at most) quadratic Hamiltonian. 
Let us hence consider the evolution corresponding to such unitary operators, which may be written as $\hat{S} = e^{i \hat{H}t}$, 
where the generating Hamiltonian $\hat{H}$ is given in Eq. (\ref{eq:quadratic}) and $t$ is a real variable representing time. 
For simplicity, let us set ${\bf r}_H =0$ to begin with. 
The Heisenberg evolution of the canonical operators promptly leads to the following linear equation:
\begin{align}
\dot{\hat{r}}_j = i[\hat{H},\hat{r}_j] &= \frac{i}{2} \sum_{kl} [\hat{r}_kH_{kl}\hat{r}_l,\hat{r}_j]  \nonumber \\
                       &= \frac{i}{2} \sum_{kl}H_{kl} \left( \hat{r}_{k}[\hat{r}_l,\hat{r}_j]+ [\hat{r}_k,\hat{r}_j]\hat{r}_l \right) 
                          =   \sum_{kl} \Omega_{jk}H_{kl} \hat{r}_l \; ,
\end{align}
which can be recast in vector form as 
\be\label{heis}
\dot{\hat{{\bf r}}} = \Omega H \hat{\bf r} \; ,
\ee
The solution of Eq.~(\ref{heis}) given the initial condition $\hat{\bf r}(0)$ is simply obtained by matrix exponentiation:
\be
\hat{\bf r}(t) = \hat{S}^\dagger \hat{\bf r}(0) \hat{S} = e^{\Omega H t} \hat{\bf r}(0).
\ee
Since they correspond to unitary transformations in the Hilbert space, transformations like $S=e^{\Omega H t}$ must preserve the 
canonical commutation relations expressed by Eq.~(\ref{ccr}), and hence the matrix $\Omega$ by congruence:
in fact, a matrix $S=e^{\Omega H t}$, with $H$ real and symmetric, belongs to the group of linear canonical transformations, 
known as the real symplectic group, satisfying the equation $S\Omega S^{\sf T} = \Omega$ 
(hence the term ``symplectic form'' for the latter). 
As an aside, let us remark that the linearity of the time evolution of the operator vector $\hat{\bf r}$ is 
the reason why continuous variable systems governed by at most quadratic Hamiltonians 
are often referred to as ``linear'' quantum system.

Under a symplectic transformation, 
first and second moments of a quantum state evolve according to the following equations
\begin{align}
{\bf r}^\prime &\rightarrow S {\bf r}^\prime \:, \\
\sig &\rightarrow S \sig S^{\sf T} \: .
\end{align}
Notice that, if the initial state is Gaussian, the two equations above completely characterise its evolution, 
since the Gaussian character of the state is preserved under quadratic Hamiltonians.

If we instead consider the case of null Hamiltonian matrix $H=0$ and non-zero linear vector ${\bf r}_H$, the unitary operator corresponds to a Weyl operator of Eq.~(\ref{eq:weyl}) -- 
with displacement vector equal to, say, ${\bf r}_t$ -- and the 
associated Heisenberg evolution of the canonical operators reads
\be
\hat{\bf r}(t) = \hat{D}_{{\bf r}_t}^{\dagger} \hat{\bf r}(0) \hat{D}_{{\bf r}_t} = \hat{\bf r}(0) + {\bf r}_t \:.
\ee
Since they are generated by Hamiltonians of order one in the canonical operators, Weyl operators send Gaussian states into Gaussian states too, resulting in the following transformations of first-moment vectors and covariance matrices:
\be
{\bf r}^\prime \rightarrow {\bf r}^\prime + {\bf r}_t \: , \quad  
\sig \rightarrow \sig \: .
\ee

Notice that one can then consider the case where both $H\neq 0$ and ${\bf r}_H \ne 0$, 
thus obtaining the most general unitary evolution preserving the Gaussian character of a quantum state. 
In general, however, such a unitary is always equivalent to the action of a purely quadratic Hamiltonian 
followed by a displacement operator.

\subsection{\bf \textsf{General-dyne measurements}} \label{s:generaldyne}

The celebrated ``coherent states'' -- ironically the ``most classical'' quantum states of the quantum optical tradition, in spite of a 
terminology that clearly bears no reference to the notion of {\em quantum} coherence -- are a particular class of Gaussian states that are 
eigenvectors of the annihilation operators $a_j=(\hat{x}+i\hat{p})/\sqrt2$. In our convention, their covariance matrix is always the identity matrix, 
while their first moments may vary arbitrarily, and determine the eigenvalue of $a_j$ they are associated with. 
In fact, any coherent state may be written as $\hat{D}_{-\bf r}\ket{0}$, that is as the action of a displacement operator 
on the vacuum state vector $\ket{0}$ (the minus sign in the displacement parameter has been inserted in order to comply with the 
standard quantum optical convention). 
It is well known, since the seminal work of Glauber \cite{Glauber}, that 
the coherent states form a resolution of the identity operator which, for $n$ modes, 
reads\footnote{For $n=1$, this equation is equivalent to the customary rendition
$$
\frac{1}{\pi}\int_{{\mathbbm C}} \ket{\alpha}\bra{\alpha} {\rm d}^{2}\alpha = \id \; ,
$$
where $\ket{\alpha}$ is the eigenvector of the annihilation operator with eigenvalue $\alpha$ $\in$ ${\mathbbm C}$. 
The complex notation above is more common in quantum optics. Eq.~(\ref{resu}) is merely given by the tensor product of $n$ 
of these identities.}
\be \label{resu}
\frac{1}{(2\pi)^n} \int_{{\mathbbm R}^{2n}} {\rm d}^{2n}r \, \hat{D}_{-\bf r} \ket{0}\bra{0} \hat{D}_{\bf r} = \id \; .
\ee
Within the framework of quantum mechanics, 
this equation implies that the set of projections on coherent states $\hat{D}_{-\bf r} \ket{0}\bra{0} \hat{D}_{\bf r}$ 
is associated with a positive operator valued measure (POVM), that is, concretely, with a physical measurement scheme.\footnote{A POVM is defined as a set of operators $\{K_{\mu}\}$ such that 
$\sum_{\mu}K_{\mu}^{\dag}K_{\mu}=\id$, where the summation over the generic label $\mu$ may generalise to an integral over one or more continuous variables. 
In the case at issue, $\ket{0}\bra{0}\hat{D}_{\bf r}$ corresponds to $K_{\mu}$, under the measure $\frac{{\rm d}^{2n}r}{(2\pi)^n}$ over ${\bf r}\in{\mathbbm R}^{2n}$.}
Such a measurement is the well known heterodyne detection scheme, which is customarily 
implemented in quantum optical laboratories.

The resolution of the identity (\ref{resu}) can be generalised by acting on both sides with a unitary transformation, that obviously preserves the identity operator. If such a unitary is a purely quadratic unitary transformation $\hat{S}$, 
corresponding to the symplectic transformation $S$, one has:
\begin{align}
\frac{1}{(2\pi)^n} \int_{{\mathbbm R}^{2n}} {\rm d}^{2n}r \, \hat{S} \hat{D}_{-\bf r} \ket{0}\bra{0} \hat{D}_{\bf r} \hat{S}^{\dag} &=
\frac{1}{(2\pi)^n} \int_{{\mathbbm R}^{2n}} {\rm d}^{2n}r \, \hat{D}_{-S{\bf r}}  \hat{S} \ket{0}\bra{0} \hat{S}^{\dag} \hat{D}_{S{\bf r}}  \label{eq:resolution} \\  
&=\frac{1}{(2\pi)^n} \int_{{\mathbbm R}^{2n}} {\rm d}^{2n}r \, \hat{D}_{-{\bf r}}  \hat{S} \ket{0}\bra{0} \hat{S}^{\dag} \hat{D}_{{\bf r}}  =  \id \; , \nonumber
\end{align}
where we used the action of a purely quadratic operation on a displacement operator: 
$\hat{S} \hat{D}_{{\bf r}} \hat{S}^{\dag} = \hat{D}_{S{\bf r}}$, 
changed the integration variables to $S {\bf r}$ and took advantage of the fact that ${\rm det}S=1$ for all $S\in Sp_{2n,{\mathbbm R}}$. 

The measurement processes described by these resolutions of the identity correspond, 
if the measurement outcome is recorded, to projections on the completely generic pure Gaussian state 
$\hat{D}_{-{\bf r}}  \hat{S} \ket{0}$. Such measurements go under the name of ``general-dyne'' measurements \cite{WisemanMilburn}, 
as they include, as we have seen, the heterodyne detection scheme for $\hat{S}=\id$ (projection on coherent states), 
and can approach arbitrarily well the homodyne detection scheme (projection on canonical operators eigenstates) 
in the limit where $\hat{S}$ is a squeezing operator with infinite squeezing parameter:
$S={\rm diag}(z,1/z)$ for $z\rightarrow\infty$. In this limit, the uncertainty on one of the canonical operator diverges while the conjugate one 
vanishes, and it can be shown that the state on which the system is projected upon is a canonical operator 
eigenstate \cite{generaldino,chiawiseman}. A deep investigation into the properties of general Gaussian 
quantum measurements (and operations) may be found in \cite{kiukas}.

In the following, we will show how to derive the evolution of Gaussian states -- described in terms 
of first and second moments -- due to the general-dyne measurement of a portion of the system 
({\em i.e.}, on part of the $n$ bosonic modes), corresponding to a projection on pure
Gaussian states.

\subsection{\bf \textsf{Conditional Gaussian dynamics}}
As we saw above, the projection on pure Gaussian states with the same second moments and varying first moments describes 
legitimate measurement processes. 
If the measurement outcome, labelled above by ${\bf r}$, is recorded, such measurements give rise to specific Gaussian CP-maps,
which can be interpreted as the filtering of the system conditioned on recording the measurement outcome ${\bf r}$.
Let us now determine how the CM of a Gaussian state is affected when a portion of the system modes is measured 
through general-dyne detection. Later on, we shall apply these formulae to derive the conditional evolution of 
continuously monitored Gaussian systems.

Given the initial Gaussian state of a system partitioned in subsystem $A$ and $B$, with CM 
$$
\sig = \left(\begin{array}{cc}
\sig_A & \sig_{AB} \\
\sig_{AB}^{\sf T} & \sig_B 
\end{array}\right) \;
$$ 
and first moments 
$$
{\bf r}' = \left(\begin{array}{c}
{\bf r}_{A}'\\
{\bf r}_{B}'
\end{array}\right) \; ,
$$
let us then determine both the probability $p({\bf r}_m)$ of measuring the general-dyne outcome ${\bf r}_m$ 
on the $m$-mode subsystem $B$ as well as the final CM and first moments of the $n$-mode subsystem $A$ given such an outcome. 
We need to evaluate the overlap between the initial state $\varrho$ and the pure Gaussian state 
on subsystem $B$ $\ket{\psi_G}_{B}$, with CM $\sig_{m}$ and first moments ${\bf r}_m$. Note that, while  
${\bf r}_m$ labels the outcome of the measurement, the CM $\sig_{m}$ characterises the specific choice of general-dyne detection.

By using the Fourier-Weyl relation (\ref{multifw}), noticing that $\bra{\psi_G}\hat{D}_{{\bf r}_{B}}\ket{\psi_G}$ is nothing but 
the characteristic function of $\ket{\psi_G}\bra{\psi_G}$ and applying the multivariate Gaussian integral (\ref{gaunt}), one gets
\begin{align}
\bra{\psi_G} \varrho \ket{\psi_G} =\,& \frac{1}{(2\pi)^{m+n}} \int_{{\mathbbm R}^{2(m+n)}} 
\hspace*{-0.9cm}
{\rm e}^{-\frac14 {\bf r}^{\sf T}\Omega^{\sf T} \sig \Omega{\bf r}+i{\bf r}^{\sf T}\Omega^{\sf T} {\bf r}'} \bra{\psi_G}\hat{D}_{\bf r}\ket{\psi_G}
\, {\rm d}{\bf r} \nonumber \\
=\,& \frac{1}{(2\pi)^{m+n}} \int_{{\mathbbm R}^{2(m+n)}} \hspace*{-0.9cm}
{\rm e}^{-\frac14 {\bf r}^{\sf T} \sig {\bf r}+i{\bf r}^{\sf T} {\bf r}'}  \hat{D}_{\Omega^{\sf T}{\bf r}_A} {\rm e}^{-\frac14 {\bf r}_{B}^{\sf T}\sig_m
{\bf r}_B-i{\bf r}^{\sf T}_B{\bf r}_m} \, {\rm d}{\bf r} \nonumber \\
=\,& \frac{2^m {\rm e}^{-({\bf r}_m-{\bf r}'_B)^{\sf T}\frac{1}{\sig_{B}+\sig_m}({\bf r}_m-{\bf r}'_B)}}{(2\pi)^{n}\sqrt{{\rm Det}(\sig_B+\sig_m)}} \;\times \label{ll} \\
&\int_{{\mathbbm R}^{2n}} \hspace*{-.4cm} {\rm e}^{-\frac14 {\bf r}_A^{\sf T} \left( \sig_A -\sig_{AB} \frac{1}{\sig_{B}+\sig_m} \sig_{AB}^{\sf T} \right) {\bf r}_A} 
{\rm e}^{i{\bf r}_A^{\sf T} 
\left({\bf r}_A'+\sig_{AB} \frac{1}{\sig_{B}+\sig_m}({\bf r}_m-{\bf r}'_B)\right)}  
\hat{D}_{\Omega^{\sf T}{\bf r}_A}  \, {\rm d}{\bf r}_A \; , \nonumber
\end{align}
which shows that, under general-dyne measurement of a set of modes, the initial 
CM $\sig_A$ and first moments ${\bf r}'_A$ of the subsystem which is not measured are mapped according to \cite{giedke02}
\begin{align}
\sig_A &\mapsto \sig_A - \sig_{AB} \frac{1}{\sig_B+\sig_m}\sig_{AB}^{\sf T} \, , \label{siggd}\\
{\bf r}'_A &\mapsto {\bf r}'_A + \sig_{AB}\frac{1}{\sig_B + \sig_m} ({\bf r}_m-{\bf r}'_B) \label{fmgd} \, ,
\end{align}
with probability density (in ${\rm d}{\bf r}_m$)
\be
p({\bf r}_m) = \frac{{\rm e}^{-({\bf r}_{m}-{\bf r}_{B}')^{\sf T} 
\frac{1}{\sig_m+\sig_B} ({\bf r}_{m}-{\bf r}_{B}')}}{\pi^m\sqrt{{\rm Det}(\sig_B + \sig_m)}} \, . \label{probgd} 
\ee
The probability above was determined by comparing the last line of Eq.~(\ref{ll}) with 
the normalisation factor of the Fourier-Weyl relation (\ref{multifw}).

At the risk of being tedious, let us remind that $\sig_{B}$ is the initial CM of the measured subset of modes, $\sig_{AB}$ contains the correlations between subsystem
of interest and measured subsystem, while $\sig_m$ is the CM of a pure Gaussian state of $n$ modes, that is a $2n\times2n$ real matrix with
determinant equal to $1$ and satisfying Inequality (\ref{eq:uncertainty}), which characterises the choice of general-dyne measurement. 
In the next section, we shall relax the requirement of purity -- ${\rm Det}\sig_{m}=1$ -- thus introducing noisy measurements, described by covariance matrices $\sig_m$ not necessarily corresponding to pure Gaussian states.
We also remark that, obviously, if no correlations are present ({\em i.e.}, if $\sig_{AB}=0$), the map above reduces to the identity, in that measuring subsystem $B$ cannot have any effect on subsystem $A$ if the two subsystems are not initially correlated.

\section{\bf \textsf{Deterministic Gaussian CP-maps}} \label{s:GaussianCPmap}

The open dynamics resulting from considering an ancillary system -- an `environment' -- in an initial Gaussian state, coupling such an ancilla to the system of interest 
through a quadratic Hamiltonian, and finally tracing out the ancilla preserves the Gaussian character of the initial state. 
Since such dynamics do not involve the probabilistic element associated with the outcome of a measurement, 
we shall refer to them as deterministic Gaussian completely positive (CP) maps. Such maps are also referred to as Gaussian ``trace-preserving'' maps.

The set of deterministic Gaussian CP-maps had already been characterised in the algebraic framework, well 
before the advent of continuous variable quantum information \cite{demoen}. It has, in more recent years, drawn considerable attention 
and has been analysed in great detail \cite{lindblad,WolfEisertCP,caruso08}. Here, we will content ourselves with deriving the main properties of such maps, 
emphasising the status of the {\em dual} maps, which will be relevant to the description of noisy measurements. 
In doing so, we will also present a particularly simple proof 
for the most general form of a deterministic Gaussian CP-map 
that has, to our knowledge, never been published before.

Given an initial Gaussian state with covariance matrix $\sig$, 
the evolution due to a deterministic Gaussian CP-map 
is completely described by two $2n\times 2n$ real matrices $X$ and $Y$, which act as follows
on first and second moments:
\begin{align}
{\bf r}^\prime &\mapsto X {\bf r}^\prime\; , \label{eq:CPmap1st} \\
\sig &\mapsto X \sig X^{\sf T} + Y  \; . \label{eq:GaussianCPmap}
\end{align}
The matrices $X$ and $Y$ must be such that
\be \label{heisdyn}
Y + i \Omega \ge i X \Omega X^{\sf T} \; .
\ee
The previous equation ensures that enough additive noise, represented by $Y$, is acting 
for the final state to satisfy the uncertainty relation (\ref{eq:uncertainty}). 
Conversely, any $X$ and $Y$ satisfying the inequality (\ref{heisdyn}) correspond to an open Gaussian dynamics as detailed above.

Showing that any open Gaussian dynamics derived from a quadratic interaction and partial tracing over a Gaussian environment 
results into a map of the form (\ref{eq:GaussianCPmap}) is straightforward and revealing.
Let us assume that a system of $n$ bosonic modes interacts with $m$ environmental modes. 
The symplectic matrix describing the joint evolution of system and environment may be split into four sub matrices, two of which, $A$ and $D$, 
describe the internal evolution of system and environment, while the other two, $B$ and $C$, issue from the quadratic coupling between 
system and environment:
\be
S = \left(\begin{array}{cc}
A & B \\
C & D 
\end{array}\right).
\ee
Notice that the most general quadratic coupling is contained in this description.
Since $S$ is symplectic, the sub-matrices $A$, $B,$ $C$ and $D$ satisfy the following matrix equality
\be
S\Omega S^{\sf T} =
\left(\begin{array}{cc}
A \Omega_n A^{\sf T} + B \Omega_m B^{\sf T} & A \Omega_n C^{\sf T} + B \Omega_m D^{\sf T} \\
C \Omega_n A^{\sf T} + D \Omega_m B^{\sf T} & C \Omega_n C^{\sf T} + D \Omega_m D^{\sf T}
\end{array}\right)
=
\left(\begin{array}{cc}
\Omega_n & 0 \\
0 & \Omega_m 
\end{array}\right) \; , \label{symplect}
\ee
where here $\Omega=\Omega_n \oplus \Omega_m$ and $\Omega_k$ is a symplectic form of $k$ degrees of freedom.

The action of the CP-map is obtained by tracing out the environmental degrees of freedom after the action of $S$ on the global state, {\em i.e.}
by considering the diagonal block of 
$S (\sig\oplus\sig_E) S^{\sf T}$ pertaining to the system, where $\sig_E$ is the initial CM of the environment. 
The matrix $\sig_E$ is only constrained by the physicality condition (\ref{eq:uncertainty}).
This evaluation yields the
evolution of the covariance matrix $\sig$ as in Eq. (\ref{eq:GaussianCPmap}), with 
\be
X=A \quad {\rm and} \quad  
Y=B \sig_{E}B^{\sf T} \; .  \label{XYAB}
\ee
The uncertainty principle (\ref{eq:uncertainty}) on the CM of the environment, 
$\sig_E+i\Omega_{m} \ge 0$, implies, for any $n\times m$ matrix $B$,
\be
B\sig_E B^{\sf T}+i B\Omega_{m} B^{\sf T} \ge 0 \; .
\ee

Because of Eq.~(\ref{symplect}) above, 
one has $B \Omega_m B^{\sf T} = \Omega_n - A \Omega_n A^{\sf T}$, which can be inserted in the previous expression to get
\be
B\sig_E B^{\sf T}+i \Omega_n - iA \Omega_n A^{\sf T} \ge 0 \; ,
\ee
which is indeed identical to the relationship (\ref{heisdyn}) between matrices $X$ and $Y$, thus completing the proof. 

The converse statement that any pair of matrices fulfilling the Inequality (\ref{heisdyn}) corresponds to a deterministic Gaussian 
CP-map as defined above is slightly more subtle to prove true. 
We report it in appendix \ref{appa} for the sake of completeness, and since we are not aware of a similarly simple proof 
of this statement to be found anywhere in the literature.
The reader who is not interested in mathematical details may skip such a demonstration, 
as nothing that follows will hinge on it. 

\subsection{\bf \textsf{Dual CP-maps and noisy measurements}}

In order to describe imperfect and noisy measurements within the Gaussian framework, 
it is expedient to introduce the notion of the dual $\Phi^*$ of a Gaussian CP-map $\Phi$.
Given $\Phi$, its dual CP-map $\Phi^*$ is defined by the following relation: 
\be
{\rm Tr} \left[\varrho_1 \Phi^*(\varrho_2)\right] = {\rm Tr} \left[\Phi(\varrho_1)\varrho_2\right] \; , 
\ee 
for all bounded operators $\varrho_1$ and $\varrho_2$.

The dual of a trace preserving Gaussian CP-map is also a Gaussian superoperator -- in the sense that it preserves the Gaussian character of the input  
characteristic function -- but is not necessarily trace preserving (it is however unital,\footnote{A unital map is one that preserves the identity operator.} as always the case for the dual of a trace-preserving map).
For a CP-map with invertible $X$, the dual map is characterised by the following $X^{*}$ and $Y^{*}$, as shown in appendix \ref{app:dual}:
\begin{align}
X^{*} &= X^{-1} \; , \label{dual1}\\
Y^{*} &= X^{-1} Y X^{-1 \sf T} \; \label{dual2} .  
\end{align}

The connection between dual CP-maps and measurements becomes clear by noticing that the Gaussian POVM 
stemming from the resolution of the identity (\ref{resu}) can be generalised by applying a unital CP-map 
on the left and right hand sides of the equation.
Since the dual of any Gaussian CP-map, determined in Eqs.~(\ref{dual1}-\ref{dual2}), is Gaussian and unital, its action 
on the POVM elements will result in what we shall refer to as a `noisy general-dyne' measurement. 
Because of the definition of a dual map, the noise described by this class of measurements is equivalent 
to applying a deterministic Gaussian CP-map on the system state before carrying out a heterodyne measurement. 
Ideal general-dyne measurements correspond to the choice $X=S^{-1}$ and $Y=0$, with $S$ symplectic, such that $X^{*}=S$ and 
$Y^{*}=0$. 
In particular, homodyne detection, corresponding to projective measurements of canonical quadratures, is retrieved as $S$ 
approaches infinite squeezing.
Beyond the unitary (symplectic) case, the noise that can be modelled 
in this class of measurements includes non-unit detection efficiency,
as well as fuzzy quadrature measurements weighted by a Gaussian mask.
Notice that simpler coarse-grained measurements of quadrature operators -- where the detection scheme simply delivers the same outcome 
for a certain interval of values of the quadrature -- do not preserve the Gaussian character of the state and are not included in our 
treatment.

More explicitly, the measurements we are considering are always equivalent to acting on the system with a Gaussian CP-map $\Phi$
before enacting the general-dyne measurement described as the projection on the pure Gaussian state $|\psi_G\rangle$, 
characterised by a covariance matrix $\sig_m$ and first moments ${\bf r}_m$. The corresponding probability reads
\be
p({\bf r}_m) = \tr{ \Phi(\varrho_G) |\psi_G\rangle\langle\psi_G|} = \tr{ \varrho_G \Phi^* (|\psi_G\rangle\langle \psi_G|)} \:.
\ee  
We can thus move the effect of the evolution on the measurement process by means of the dual-map $\Phi^*$ -- notice that the unitality is necessary and sufficient in order to preserve the resolution of the identity (\ref{eq:resolution}). As a consequence, the measurement is described by Gaussian operators characterised by the covariance matrix
\be
\sig_m^* = X^* \sig_m X^{* {\sf T}} + Y^* \:. \label{eq:sigmdual} 
\ee
Any physical CM $\sig$, such that $\sig+i\Omega\ge0$, may be obtained as the action of a dual CP map on the CM 
corresponding to a pure state.\footnote{This a straightforward consequence of the normal mode decomposition of a 
CM: any physical CM $\sig$ is strictly positive (as a consequence of $\sig+i\Omega\ge 0$), and hence a symplectic transformation $S$ 
exists such that $\sig =S {\boldsymbol \nu} S^{\sf T}$, where ${\boldsymbol \nu}=\bigoplus_{j=1}^{n}\nu_j \id_2$. The real quantities 
$\nu_j$ are called symplectic eigenvalues of the CM $\sig$. 
Because of the uncertainty relation, the symplectic eigenvalues satisfy $\nu_j\ge 1$ $\forall$ 
$j$, and a Gaussian state is pure if and only if the symplectic eigenvalues of its CM are all equal to $1$. Then, a Gaussian state with 
CM $\sig=S {\boldsymbol \nu} S^{\sf T}$ is the output of a Gaussian CP-map with $X=\id$ and $Y = S({\boldsymbol \nu}-\id)S^{\sf T}$ 
under the pure input with CM $S S^{\sf T}$.} Therefore, the probability outcome 
of the most general noisy general-dyne detection may be evaluated 
as ${\rm Tr}\left[\varrho \varrho_G\right]$, where $\varrho$ is the state of the system to be measured and $\varrho_G$ is the 
most general, possibly mixed, Gaussian state.

It is easy to show that, even if $\sig_m^*$ does not in general correspond to the CM of a pure state,  
the conditional states and the measurement probability are still obtained by replacing $\sig_m$ with $\sig_m^*$ in Eqs. (\ref{siggd}), (\ref{fmgd}) and (\ref{probgd}).

\section{\bf \textsf{Open diffusive dynamics}} \label{s:openGaussian}
In the following, we shall consider a system weakly coupled to a large environment, 
whose correlation times are much shorter than the system dynamical time-scales, such that no information leaking to the 
environment is ever fed back into the system. A bath of this type is usually referred to as a memoryless, or Markovian, bath, 
while the whole set of dynamical conditions we will consider, including that of weak coupling, that basically does not alter the state of the bath, are collectively known as the Born-Markov regime.

More specifically, we will assume white quantum noise in the so-called input-output formalism, which describes a system 
of $n$ modes 
in contact with a continuous train of $m$ incoming bosonic modes $\hat{\bf r}_{in}(t)$, 
each of which interacts with the system at time $t$ and is then
scattered as the output mode $\hat{\bf r}_{out}(t)$. Note that, when associated to input and output fields,  
the parameter $t$ is not a dynamical variable but a label to distinguish the input modes that interact with the system at time $t$. 

Let us first state the ``white noise'' condition, that entails a Markovian dynamics:
\be
\langle \{\hat{\bf r}_{in}(t),\hat{\bf r}_{in}^{\sf T}(t')\} \rangle = \sig_B \, \delta(t-t') \; , \quad \sig_B+i\Omega \ge 0 \label{whn} \, .
\ee
In other words, the system interacts at each instant $t$ with a different set of modes, completely uncorrelated with those  
it encountered in the past. We allow for complete generality in the second order correlations of the bath quadratures, by 
letting their covariance matrix $\sig_B$ be any physical CM. Note that this allows one to set a finite environmental temperature. 

The white noise condition (\ref{whn}) may be recast as a condition on infinitesimal quantum operators, 
which act as counterparts of classical stochastic increments in what is known as quantum stochastic calculus. 
Forfeiting a rigorous mathematical framework, which would unnecessarily burden our treatment, 
let us define the operators $\delta \hat{\bf r}_{in}(t)= \int_t^{t+\delta t}\hat{\bf r}_{in}(s){\rm d}s$
for a certain arbitrary interval $\delta t$. Notice that: 
\be
\langle \{\delta\hat{\bf r}_{in}(t),{\delta}\hat{\bf r}_{in}^{\sf T}(t)\} \rangle 
=\sig_B \int_{t}^{t+\delta t}{\rm d}s = \sig_B \delta t \label{wiener} \; .
\ee
Hence, in the limit of arbitrarily small ${\delta t}={\rm d}t$, one may define the infinitesimal bath quadrature operators 
${\rm d}\hat{\bf r}_{in}(t) = \hat{\bf r}_{in} {\rm d}t$ and obtain
\be
\langle \{{\rm d}\hat{\bf r}_{in}(t),{\rm d}\hat{\bf r}_{in}^{\sf T}(t')\} \rangle = \sig_B {\rm d}t \label{wiener} \; .
\ee
If we make the assumption that the bath is in a Gaussian state, which we shall, 
Eq.~(\ref{wiener}) defines the `quantum Wiener process' ${\rm d}\hat{\bf r}_{in}$, 
which we inferred through a heuristic argument from the white noise condition (\ref{whn}) 
but could have otherwise just been postulated.
Our argument bridges between 
the open quantum system approach in the Schr\"odinger picture, where the emphasis is often on the spectral properties of the reservoir,  
and quantum stochastic calculus.

Without dwelling on formal definitions, let us proceed to define the operator 
$\hat{\bf r}'_{in}$ as per $\hat{\bf r}_{in}{\rm d}t = \hat{\bf r}'_{in} {\rm d}w$, and then state the so called Ito rule 
\be
{\rm d}w^2 = {\rm d}t \; . \label{ito}
\ee 
Notice that, while compliant with the expectation values of Eq.~(\ref{wiener}), Eq.~(\ref{ito}) is stronger, in that it holds 
{\em deterministically} (and not only on average) for the Wiener process ${\rm d}w$ \cite{JacobsCP}. 
The fact that the squared increment is proportional to the infinitesimal time interval is a general property of
Wiener processes that is common to all continuous (but not differentiable) diffusive dynamics, such as a 
continuous random walk, where the variance of the process grows linearly in time.
We shall make use of this non-trivial statement later on, in the derivation of conditional dynamics 
due to the measurement of $\hat{\bf r}'_{in}$. The latter is an array of canonical operators corresponding to a discrete set of bosonic modes 
-- in the sense that it fulfils Eq.~(\ref{ccr}) -- and may be thought of as the set of photon wave-packets (discrete travelling modes) that undergo detection.

Let us now introduce a quadratic coupling Hamiltonian $\hat{H}_{C}$ 
between system and input modes (bath):
\be
\hat{H}_{C} = \hat{\bf r}^{\sf T} C \hat{\bf r}_{in} = \frac12 \hat{\bf r}^{\sf T}_{sb} H_{C} \hat{\bf r}_{sb} =  
\frac12\hat{\bf r}^{\sf T}_{sb} \left(\begin{array}{cc}
0 & C \\
C^{\sf T} & 0
\end{array}\right) \hat{\bf r}_{sb} \, , \label{eq:Hc}
\ee  
where the real $2n\times2m$ coupling matrix $C$ is entirely generic\footnote{It is possible to show that, given 
any coupling matrix $C$, a quadratic Hamiltonian for the continuous set of modes of the bath may always be found such that the
global Hamiltonian is a positive operator, as required by thermodynamic stability for any physical system. 
Hence, any coupling matrix $C$ corresponds to a {\em bonafide} physical evolution. Notice that here we do not address the problem of which 
couplings are allowed by stability once the Hamiltonian of system and bath are fixed.} 
(recall that any mode of the system could interact with any number of 
input modes) and $\hat{\bf r}^{\sf T}_{sb} = (\hat{\bf r}^{\sf T},\hat{\bf r}_{in}^{\sf T})$. 
Notice now that, under such a coupling, 
the dynamics of the quantum variables $\hat{\bf r}^{\prime}_{sb} = (\hat{\bf r},\hat{\bf r}'_{in})$
over an interval ${\rm d}t$ 
is generated by the operator $\hat{\bf r}^{\sf T}_{sb} H_{C} \hat{\bf r}_{sb}\,{\rm d}t  = 
\hat{\bf r}^{\prime \sf T}_{sb} H_{C} \hat{\bf r}'_{sb}{{\rm d}w}$. Therefore, 
their initial CM $\sig\oplus\sig_B$ evolves under the symplectic transformation 
\be
{\rm e}^{\Omega H_{C}{{\rm d}w}} = \id + \Omega{H}_{C}{{\rm d}w} + \frac{(\Omega{H_{C}})^{2}}{2}\,{\rm d}t + o({\rm d}t) \, ,
\ee
which acts by congruence as follows
\begin{align}
{\rm e}^{\Omega{H}_{C} {{\rm d}w}} \left(\sig\oplus\sig_B\right) {\rm e}^{(\Omega{H_{C}})^{\sf T} {{\rm d}w}} 
 =&
\left(\sig\oplus\sig_B\right) + \left(\frac{\Omega C \Omega{C^{\sf T}}\sig+\sig C\Omega C^{\sf T}\Omega}{2}\right)\oplus \tilde{\sig}_{B,1} \,{\rm d}t 
\nonumber \\
& + \Omega{C} \sig_B {C}^{\sf T}\Omega^{\sf T}\oplus \tilde{\sig}_{B,2} \, {\rm d}t + \sig_{SB} {{\rm d}w} + \sig_{BB}
\, o({\rm d}t) \; , \label{diffu} 
\end{align}
where the Ito rule ${\rm d}w^2={\rm d}t$ was applied, the Landau symbol $o({\rm d}t)$ employed, and
\begin{align}
\sig_{SB} &= \left(\begin{array}{cc} 
0 & \Omega{C} \sig_{B} + \sig {C\Omega^{\sf T}} \\
\sig_{B}{C}^{\sf T}\Omega^{\sf T} + \Omega{C^{\sf T}} \sig & 0
\end{array}\right) \:,\\
\tilde{\sig}_{B,1} &=  \frac{\Omega C^{\sf T} \Omega C \sig_B+\sig_B C^{\sf T}\Omega C \Omega}{2} \:, \\
\tilde{\sig}_{B,2} &= \Omega^{\sf T} C^{\sf T} \sig C \Omega \:,
\end{align}
while $\sig_{BB}$ is matrix with support only on the bath variables (and thus irrelevant as to the evolution of the system, since the bath state is 
refreshed at every instant under the white noise assumption). 

The unconditional dynamics of the system occurring when the bath is disregarded 
-- or, equivalently, when hypothetical measurements on the bath are not recorded --
is obtained by considering only the diagonal block pertaining to the system in the matrix equation (\ref{diffu}), 
and takes the form of a diffusion equation:
\begin{align}
\dot{\sig} &= A \sig + \sig A^{\sf T} + D \; ,  \label{diffusion}
\end{align}
for the following drift and diffusion matrices $A$ and $D$:
\begin{align}
A = \frac{\Omega C \Omega{C^{\sf T}}}{2} \; , \quad  
D &= \Omega{C} \sig_B {C}^{\sf T}\Omega^{\sf T} \; . 
\end{align}
If one assumes that the expectation value of the canonical operators of the bath vanishes 
(${\rm Tr}\left[\varrho_B \hat{\bf r}_{in}\right]=0$), then
the unconditional evolution equation of the first moments vector ${\bf r}'$ is simply given by
\be
\dot{\bf r}' = A {\bf r}' \; . \label{diffusionFM}
\ee
Situations where the input fields have non-zero first moments describe `driven' systems,
where a vector independent from the system state is added to right hand side of the equation above. 
For simplicity, we will not consider driving in the following although it can be easily accommodated in the picture.

We should also remark that, in order to keep the derivation cleaner, we have not considered 
any Hamiltonian operator describing the dynamics of the system alone. If such a Hamiltonian, 
quadratic in the system canonical operators, is included as
\be
\hat{H}_s = \frac12 \hat{\bf r}^{\sf T} H_s \hat{\bf r} \:, 
\ee
then only the drift matrix is modified and takes the form 
\be
A = \Omega H_s +  \frac{\Omega C \Omega{C^{\sf T}}}{2}\:. \label{eq:driftHamilton}
\ee

It is interesting to consider briefly which unconditional diffusive evolutions are allowed within this framework, that is, what are the most 
general drift and diffusion matrices $A$ and $D$. Eq.~(\ref{eq:driftHamilton}) shows that $\Omega^{\sf T} A = H_s + C\Omega C^{\sf T}/2$. 
Since $C$ is completely generic (not even necessarily square), $C \Omega C^{\sf T}$ is a completely generic $2n\times 2n$ 
real anti-symmetric matrix, while $H_s$ is a completely generic symmetric matrix. Hence, $A$ is any real square matrix, 
and the antisymmetric and symmetric parts of $\Omega^{\sf T}A$ give, respectively, $H$ and $C\Omega C^{\sf T}/2$. 
The set of allowed $D$'s given $A$ is determined as follows.
Let $A_a=C\Omega C^{\sf T}$ be the anti-symmetric part of $2\Omega^{\sf T} A$. 
Then, $D$ must only comply with the uncertainty relation of the bath state $\sig_B+i\Omega\ge 0$, which entails
$D+i\Omega C\Omega C^{\sf T}\Omega^{\sf T} = D+i\Omega A_a \Omega^{\sf T} \ge 0$. 
Summing up, one can characterise the most general $A$ and $D$ as (see also \cite{WisemanMilburn}):
\begin{align}
D + i\Omega A_a \Omega^{\sf T} & \ge 0 \; , \\
{\rm with} \quad A \in {\mathcal M}_{2n,2n}({\mathbbm R}) \quad {\rm and}& \quad A_a = \Omega^{\sf T} A - A^{\sf T} \Omega  \; .
\end{align}

For a single degree of freedom, when the matrices above are all two-dimensional, 
the condition above reduces to ${\rm Det}D \ge {\rm Det}A_a$, which encompass all single-mode unconditional diffusive dynamics.

\subsection{\bf \textsf{Master equations \label{master}}}

The diffusive dynamics we have considered above, including only a linear coupling to the bath, are entirely 
characterised by the matrices $A$ and $D$. The details of the evolution of a Gaussian state, given 
by Eqs.~(\ref{diffusion}) and (\ref{diffusionFM}), thus completely specify such dynamics. 
As a consequence, the equation of motion governing the evolution of a generic quantum state $\varrho$,
the so called master equation, 
may in principle be inferred from the Gaussian dynamics. 

In point of fact, the evolution of a Gaussian state with zero first moments, which remain zero 
as per Eq.~(\ref{diffusionFM}), is sufficient to derive the corresponding master equation. 
The time-derivative of a Gaussian characteristic function $\chi_G$ with ${\bf r}'=0$ under the diffusive evolution 
of Eq.~(\ref{diffusion}) is given by
\be
\dot{\chi_G} = -\frac14 {\bf r}^{\sf T} \Omega^{\sf T} (A\sig+\sig A^{\sf T}+D) \Omega{\bf r}\, \chi_G \; . \label{chit}
\ee
This may be rewritten as a linear differential equation for a generic characteristic function $\chi$, which already provides the general 
dynamics of any quantum state, since any quantum state allows for a description in terms of a characteristic function. 
In turn, differential operators acting on $\chi$ are equivalent to linear operators multiplying the density matrix $\varrho$, 
so that one may eventually retrieve the so called master equation, that governs the evolution of the density matrix of the open quantum system. 
For a detailed derivation of this well known equivalence see, for instance, 
Appendix 12 of \cite{barnettradmore}. 
In terms of the quadrature operators that we are utilising, the correspondence reads, for a single degree of freedom:
\begin{align}
\left(-i\partial_p -\frac{x}{2} \right)\chi_O \; &\longleftrightarrow \; \hat{x} \hat{O}  \quad , \quad 
\left(i\partial_x -\frac{p}{2} \right)\chi_O \; \longleftrightarrow \; \hat{p} \hat{O} \; , \label{chirho1} \\
\left(-i\partial_p + \frac{x}{2} \right)\chi_O \; &\longleftrightarrow \;  \hat{O} \hat{x}  \quad , \quad 
\left(i\partial_x + \frac{p}{2} \right)\chi_O \; \longleftrightarrow \; \hat{O} \hat{p} \; , \label{chirho2}
\end{align}
where $\chi_O$ stands for the characteristic function of operator $\hat{O}$.

Let us provide a concrete example of how such a derivation would proceed, in the simple but especially relevant case of a single mode 
interacting with a white noise reservoir in the vacuum state ({\em i.e.}, at zero temperature), through the rotating wave coupling. 
This situation typically models electromagnetic radiation in a lossy cavity at high frequencies and corresponds, in our compact notation, 
to $\sig_B=C=\id$, which implies $A=-\frac12\id$ and $D=\id$. Then, Eq.~(\ref{chit}) reads 
$$
\dot{\chi_G} = -\frac14 {\bf r}^{\sf T} \Omega^{\sf T} (\id-\sig) \Omega{\bf r} \, \chi_G 
$$
which, for a single mode, is equivalent to the following equation on a generic, non necessarily Gaussian, $\chi$:
\be
\dot{\chi} = -\frac14 \left( 2x\partial_x + 2p\partial_p +x^2+p^2 \right) \chi \; . \label{qomechi} 
\ee
Eq.~(\ref{qomechi}) describes the diffusive dynamics of a generic quantum state with characteristic function $\chi$. 
One may then apply Eqs.~(\ref{chirho1}-\ref{chirho2}) to obtain the (completely equivalent) master equation description 
for the density matrix $\varrho$:
\be
\dot{\varrho} = a\varrho a^{\dag} -\frac12 \left( \varrho a^\dag a + a^\dag a \varrho \right)\; , \label{qoma}
\ee
where we switched to the annihilation operator $a=(\hat{x}+i\hat{p})/\sqrt2$. 
The master equation (\ref{qoma}) is so widely applied in quantum optics that is sometimes refereed to 
as the ``quantum optical'' master equation.
It is customary to define the superoperator 
${\mathcal D}[\hat{O}]= \hat{O}\varrho \hat{O}^{\dag} -\frac12 \left( \varrho \hat{O}^\dag \hat{O} + \hat{O}^\dag \hat{O} \varrho \right)$ 
for a generic operator $\hat{O}$
and to write the master equation as $\dot{\varrho}={\mathcal D}[a]\varrho$. 
For the reader's convenience, we shall explicitly state corresponding master equations, 
as well as stochastic master equations in the monitored case, in section \ref{examples}, when 
we will be discussing specific examples of diffusive dynamics.

Our derivation of the master equation in the Born-Markov regime requires the relatively sophisticated machinery 
of the characteristic function description but, arguably, isolates the key conceptual issues more clearly than the standard 
Hilbert space derivation, usually followed in the open quantum systems literature \cite{carmichael}.

\section{\bf \textsf{General-dyne filtering of diffusive dynamics}} \label{s:filtering}

The finite off-diagonal term $\sig_{SB}{{\rm d}w}$ in
Eq.~(\ref{diffu}) shows that, at every instant in time, correlations build up between the system and the mode 
it interacted with. Hence, if the output mode corresponding to the interacting input one is measured, one can influence the system dynamics.
Let us then determine such a conditional dynamics when the measurement of the environmental mode is 
a general-dyne detection. These continuous, `weak' measurements, whereby the system is not directly observed, but only through the 
environmental modes with which it interacted for an infinitesimal interval ${\rm d}t$, are referred to as general-dyne ``monitoring''.

By applying the Ito formula (\ref{ito}) and the Eqs.~(\ref{siggd}) and (\ref{fmgd}) to the covariance matrix of Eq. (\ref{diffu}), 
with $\sig_{AB}$ replaced by the off-diagonal block of $\sig_{SB}$,
one promptly obtains the evolution equation of the monitored covariance matrix and of the first moments.
Here, $\sig_m$ parametrises the noisy general-dyne measurement of the bath degrees of freedom.

The covariance matrix obeys the following deterministic Riccati (quadratic) equation:
\begin{align}
\dot{\sig} = A\sig + \sig A^{\sf T} + D  
- ( \Omega C \sig_{B} - \sig C \Omega) \,\frac{1}{\sig_B + \sig_m}\, 
(\Omega C^{\sf T} \sig - \sig_B C^{\sf T} \Omega) \:, \label{briccone}
\end{align}
that can be rewritten as
\begin{align} 
\dot{\sig} &= \tilde{A} \sig + \sig \tilde{A}^{\sf T} + \tilde{D} - \sig B B^{\sf T} \sig \label{eq:diffusofiltering}
\end{align}
where
\begin{align}
\tilde{A} &= A - \Omega C \sig_B \frac{1}{\sig_B + \sig_{m}}\Omega C^{\sf T} \:, \\
\tilde{D} &= D + \Omega C \sig_B \frac{1}{\sig_B + \sig_{m}} \sig_B C^{\sf T} \Omega \:, \\
B &= C \Omega\, \sqrt{\frac{1}{\sig_B + \sig_{m}}} \:.
\end{align}
The deterministic nature of the second moments' evolution is a peculiar property of Gaussian measures, 
and is not related in any way to the time-continuous, noisy stochastic process the system is undergoing. In fact, 
the update of the covariance matrix of a Gaussian state in the general case, given by Eq.~(\ref{siggd}), 
is always independent from the measured outcome and hence `deterministic'.
Note also that, as apparent to Eq.~(\ref{briccone}) and as one should expect, 
general-dyne filtering always implies a reduction of noise, in the sense 
that a positive matrix is subtracted from the time-derivative of $\sig$ with respect to the unconditional, unfiltered case.

\begin{figure}[t!]
\begin{center}
\includegraphics[scale=0.5]{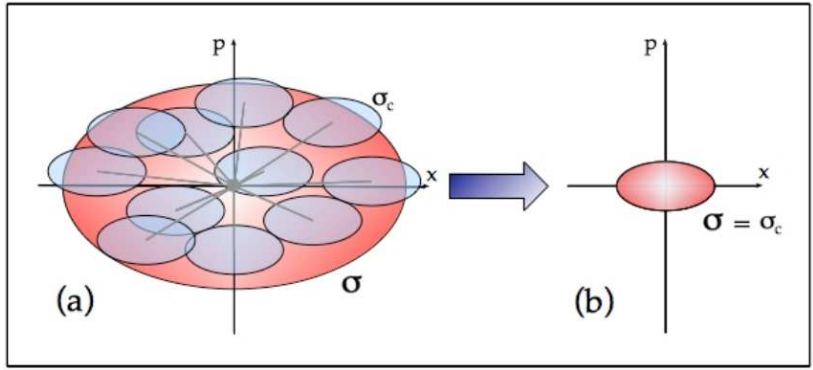}
\end{center}
\caption{\label{feedbacco}Heuristic phase-space representation of Gaussian filtering and 
possible linear feedback action on a single bosonic mode. 
(a) The covariance matrix $\sig$ of the unfiltered, unconditional state is represented by the large ellipse in the background, while 
the conditional states are represented by the smaller ellipses with the same shape (since the evolution of the second moments is 
deterministic) but varying centres, which represent the different first moments 
(as the evolution of the first moments is a stochastic Wiener process). 
(b) For each such general-dyne filtering, a Markovian linear feedback action exists such that the covariance matrix of the resulting deterministic, unconditional state is the same as the conditional covariance matrix $\sig_c$. The first moments may instead be set at one's leisure 
(they are set to zero in the picture).
The feedback is implemented through Weyl displacement operators with parameters proportional to the measurement outcomes.}
\end{figure}

The $1^{st}$ moments' conditional evolution is instead stochastic, and completely analogous to a classical Wiener process: 
\be
{\rm d}{\bf r}' = A {\bf r}' \,{\rm d}t +( \Omega C \sig_{B} - \sig C \Omega)\left(\frac{1}{\sig_B + \sig_m}\right)^{\frac12} {\rm d}{\bf w}, \label{eq:diffusofilteringFM}
\ee
with $\langle\{ {\rm d}{\bf w},{\rm d}{\bf w}^{\sf T} \}\rangle = \id \,{\rm d}t$, upon identifying ${\rm d}{\bf w}=(\sig_B+\sig_m)^{-1/2}({\bf r}_m-{\bf r}'_{B})$
in Eqs.~(\ref{fmgd}) and (\ref{probgd}).  
Notice that ${\rm d}{\bf w}$ is not quite defined as a standard Wiener increment, since in components one has $\langle {\rm d}w_j^2 \rangle = {\rm d}t/2$ (rather than just ${\rm d}t$ as customary). The measurement results, upon which the evolution of the quantum state is conditioned, 
are often expressed as a real `current' ${\bf y}$ with uncorrelated noise, defined by:
\be
{\bf y} \,{\rm d}t = - B^{\sf T} {\bf r}^\prime \,{\rm d}t + {\rm d}{\bf w}.
\ee
The noise reduction resulting from general-dyne filtering is illustrated in Fig.~\ref{feedbacco}a, where we emphasise that the 
unfiltered state is the Gaussian average of an ensemble of Gaussian states with the same second moments and varying 
first moments. It can also be shown that, given any general-dyne measurement, the first moments of the filtered state may always 
be displaced to a fixed point (chosen, for simplicity, as the origin in Fig.~\ref{feedbacco}b) by a feedback action enacted through 
controlled Weyl displacement operators \cite{WisemanDoherty}. The latter are unitary operations of the form of Eq.~(\ref{eq:weyl}), and only require Hamiltonians 
which are linear in the canonical operators, so that such feedback actions are usually referred to as ``linear feedback''.

The equations above, that completely characterise the conditional evolution of Gaussian states under general-dyne monitoring, 
are identical to the evolution equations of a classical Kalman filter in a linear, Gaussian classical system. 
Note however that our derivation was entirely based on the update of the quantum state resulting, for indirect measurements 
such as ours, from the von Neumann postulate and from the tensor product structure of composite Hilbert spaces. 
We did not invoke any other filtering criterion, which were explored in other strands of research in quantum mechanics following a tradition that 
goes back to seminal work by Belavkin \cite{belavkin92,belavkin80,belavkin92b,jamesrs1,jamesrs2,james}. 
Let us remark that the striking, and not at all trivial, analogy with classical Kalman filtering equations, 
which are based on the minimisation of a squared distance, 
may serve as a powerful pragmatic tool in achieving the real-time update of quantum systems in experiments \cite{aspel}.

This correspondence with classical filtering 
is yet another consequence of the {\em apparent} classical-like nature of quantum Gaussian systems. 
If one restricts to 
Gaussian measurements, Gaussian quantum states and dynamics may always be mimicked by classical stochastic variables, 
with the only distinctive feature of having to obey the uncertainty principle.
However, one should not forget that quantum Gaussian states do involve quantum coherence in the underlying Hilbert space description, 
a property that no classical variable may ever boast. This justifies the interest in quantum Gaussian states for quantum technologies, 
with the caveat that, at some point, coherent quantum resources will have to be harnessed through non-Gaussian means, such as 
photon-number detectors or Gaussian detectors combined with non-Gaussian operations.\footnote{With the possible exception of squeezing, 
which is interesting per se as a means to achieve unprecedented sensitivities, on scales where classical variables may not even be defined.}

The choice of the general-dyne filter was given here in terms of a Gaussian state covariance matrix 
$\sig_{m}$, with the only constraint that $\sig_m+i\Omega\ge0$. 
As shown in Sec. \ref{s:GaussianCPmap}, the extension of $\sig_m$ to covariance matrices corresponding to generic, mixed states
allows on to model detectors subject to inefficiency (loss) and Gaussian coarse-graining.

Implicitly, we have assumed in our treatment that the environment, whose state is represented by the covariance matrix $\sig_B$, 
has a certain number of degrees of freedom, say $m$, and that the continuous general-dyne measurement being performed is also
parametrised by a $2m\times 2m$ covariance matrix $\sig_m$. However, when the state of the environment is a 
mixed quantum state, one could extend such a description by replacing $\sig_B$ with the CM corresponding to any Gaussian purification of the 
bath state (whose submatrix pertaining to the original degrees of freedom of the bath is still $\sig_B$), and then consider 
any physical $\sig_m$ in the extended phase space. It turns out that such a wider class of monitoring schemes outperforms 
the ones restricted to the detection of the original environmental modes 
in the optimisation of certain figures of merits, such as steady-state squeezing and quantum entanglement 
\cite{Tempura,Ravotto}.
Such a larger class of filters may also be expediently parametrised through the so called ``unravelling matrix'', 
introduced by Wiseman and Diosi \cite{WisemanDiosi}. The term ``unravelling'' 
has a long standing tradition in the theory of open quantum systems, referring to the fact that the unconditional evolution 
described by a quantum master equation may always be ``unravelled'' into an ensemble of conditional stochastic quantum trajectories (each 
of them corresponding to a certain sequence of outcomes and state updates resulting from monitoring the environment). The average over the quantum trajectories yields back the unconditional evolution. Each possible choice of measurements performed on the environment is known as an
``unravelling'' of the master equation.

Our alternate parametrisation in terms of $\sig_B$ and $C$, besides being simpler and allowing for a more compact notation, has the advantage 
of immediately relating to a physical detection scheme in the general-dyne framework. 
The unravelling matrix parametrisation, instead, although it encompasses the same class of  
measurements, does not allow one to systematically retrieve the associated detection scheme.   
On the other hand, the unravelling matrix parametrisation enjoys certain advantages when deriving general results, such as the class of 
all stabilising solutions of the Riccati equation (\ref{briccone}) which, as shown in \cite{WisemanDoherty}, turns out to be all the $\sig$ satisfying
\be
A\sig+\sig A^{\sf T} + D \ge 0 \; , \quad \sig+i\Omega\ge0 \; .
\ee
Yet other alternative parametrisations of the general-dyne unravellings are derived and discussed \cite{chiawiseman}. 

Let us also remark that here we are considering a fixed general-dyne filter, not accounting for 
more general unravelling associated with adaptive measurements, where the choice of the measure changes in time. 
These turn out to be advantageous in certain tasks where transient dynamics are relevant, such as optical phase estimation \cite{phasest}.

\section{\bf \textsf{Examples and applications \label{examples}}} 
Let us now demonstrate the effectiveness of the formalism introduced above 
by applying it to some selected cases of practical interest.
\subsection{\bf \textsf{The quantum optical parametric oscillator}} \label{s:examples}
The mechanism of parametric amplification, through which one obtains squeezed Gaussian states
of a single electromagnetic degree of freedom, 
may be described by the Hamiltonian \cite{reid}
\be
\hat{H}_s = -\frac\chi 2 (\hat{x} \hat{p} + \hat{p} \hat{x} ) \: .
\ee
By assuming the system is interacting as per the previous section with a Markovian bath
in thermal equilibrium, described by a single-mode covariance matrix
\be
\sig_B = (2 n_{\sf th} + 1) \mathbbm{1}_2 \:,
\ee
(where $n_{\sf th}$ corresponds to the average number of thermal photons), one obtains a model for an optical parametric oscillator. 
In the quantum optics laboratory, such a device consists in an optical cavity mode interacting with a non-linear optical crystal with finite second order susceptibility and driven by an external laser \cite{OPA}. The coupling constant $\chi$ is given by the second order susceptibility times the average photon number of the driving laser. 
After adiabatic elimination of the crystal's degrees of freedom, one gets the effective Hamiltonian $\hat{H}_s$ given above.
The interaction between the cavity mode and the environment is described by a Hamiltonian $\hat{H}_C$ as in Eq. (\ref{eq:Hc}), with $C = \sqrt{\gamma} \mathbbm{1}_2$, 
such that 
\be
\hat{H}_C \:{\rm d}t = \sqrt{\gamma} ( \hat{x} \hat{x}_{in}^\prime + \hat{p} \hat{p}_{in}^\prime) dw \:, \label{eq:interactionA}
\ee
corresponding to a passive (beam-splitter) interaction between system and bath.

If the environment is not monitored, then the evolution for the Gaussian state of the system is described by Eqs. (\ref{diffusion}) and (\ref{diffusionFM}), with drift and diffusion matrix that can be easily evaluated as
\begin{align}
A &= \Omega H_s + \frac{\Omega C \Omega C^{\sf T}}{2} = 
\left(\begin{array}{c c}
-\chi - \gamma/2 & 0 \\
0 & \chi - \gamma/2 
\end{array}\right) , \\
D &= \Omega C \sig_B C^{\sf T} \Omega^{\sf T} = \gamma (2 n_{\sf th} +1 )\mathbbm{1}_2 \:. 
\end{align}

By applying the method detailed in section \ref{master}, one may check that the same evolution for first and second moments can be 
obtained by starting from the well know quantum optical master equation, describing the loss mechanism of a bosonic mode interacting with a non-zero temperature Markovian bath, {\em i.e.}
\be
\dot{\varrho} = - i [ \hat{H}_s ,\varrho ] + \gamma (n_{\sf th} + 1) \mathcal{D}[\hat{a}] \varrho
+\gamma n_{\sf th}\mathcal{D}[\hat{a}^{\dagger}] \varrho \:, \label{eq:thermalME}
\ee
in terms of the annihilation operator $\hat{a} = (\hat{x} + i \hat{p})/\sqrt{2}$ and the superoperator
$\mathcal{D}[O] \varrho = O \varrho O^{\dag} - (O^{\dag} O \varrho + \varrho O^{\dag} O)/2$.\\

This unconditional dynamics is stable, in the sense of admitting a steady state, for $\chi<\gamma/2$. In the stable region, 
the steady state covariance matrix is readily obtained by setting $\dot{\sig}=0$ in the diffusion equation, yielding
\be
\sig = \left(\begin{array}{cc} 
\frac{1}{1+\frac{2\chi}{\gamma}} & 0 \\
0 & \frac{1}{1-\frac{2\chi}{\gamma}} 
\end{array}\right) \; . \label{uncss}
\ee 
If one quantifies the squeezing by the smallest eigenvalue of $\sig$ (a good indicator of the noise on the least noisy quadrature operator, 
with lower values denoting more squeezing),
the inspection of Eq.~(\ref{uncss}) immediately reveals a level of squeezing that goes from $1$ (no squeezing) for $\chi=0$ 
(where the steady state is obviously just the vacuum) to a still finite minimum of $1/2$ at the instability point $\chi=\gamma/2$. 

We now consider the case where the system is monitored continuously via general-dyne detection. Aside from 
an additional phase-rotation, that we will ignore in the following for the sake of simplicity, a general-dyne detection corresponding to 
Eq.~(\ref{eq:resolution}) is described by projection on Gaussian states having a covariance matrix 
$$
\sig_{m} = {\rm diag}( s , 1/s) \:, \:\:\: s>0 \, .
$$
The limits $s\rightarrow 0$ and $s\rightarrow \infty$ describe, respectively, the homodyne detection of the quadrature operators $\hat{x}_{in}^\prime$ and $\hat{p}_{in}^\prime$, and thus the indirect monitoring of the system quadratures $\hat{p}$ and $\hat{x}$ (because of the interaction Hamiltonian (\ref{eq:interactionA})). The choice $s=1$ describes heterodyne detection ({\em i.e.} projection on coherent states), while measurements corresponding to intermediate values of $s$ can be easily implemented by using linear optics and homodyne detectors \cite{generaldino,chiawiseman}.

For the sake of argument, let us consider homodyne detection of the position operator $\hat{x}$ ($s\rightarrow \infty$). The evolution for the covariance matrix can be easily computed, obtaining Eq. (\ref{eq:diffusofiltering}), with matrices
\begin{align}
\tilde{A} = A + \left( 
\begin{array}{c c}
\gamma & 0 \\
0 & 0
\end{array}
\right) \; ,\quad 
\tilde{D} = D - \left( 
\begin{array}{c c}
\gamma (2 n_{\sf th} + 1) & 0 \\
0 & 0
\end{array}
\right) \; , \quad 
B =
\left( 
\begin{array}{c c}
0 & \sqrt{\frac{\gamma}{ 2 n_{\sf th} + 1}} \\
0 & 0
\end{array}
\right) \:. \nonumber
\end{align}

Since all the matrix involved are diagonal, one can straightforwardly obtain the analytical solution of 
Eq. (\ref{eq:diffusofiltering}) for the steady-state covariance matrix (obtained by setting $\dot{\sig}=0$), which reads
\begin{align}
\sig &= 
\left(
\begin{array}{c c}
\frac{(\gamma - 2\chi)(2 n_{\sf th} + 1)}{\gamma}& 0 \\
0 & \frac{\gamma(2 n_{\sf th} +1)}{\gamma - 2\chi}
\end{array}
\right) \; .
\end{align}
Setting, for simplicity, the number of thermal excitations $n_{\sf th}=0$, one observes that the steady state squeezing 
in the $x$-quadrature 
improves for all values of $\chi$ with respect to the value $\frac{1}{1+\frac{2\chi}{\gamma}}$ obtained above in the unfiltered case.  
In principle, an infinite amount of squeezing 
may be achieved through monitoring near instability, {\em i.e.} for $\chi=\gamma /2$. 
This remarkable improvement over the non-monitored, unconditional case,
is obviously due entirely to the detection the environment is undergoing. 
This Gaussian dynamics corresponds to the stochastic master equation derived from the monitoring of the quadrature $\hat{x}$ via homodyne detection, 
\be
{\rm d}{\varrho} =  - i [ \hat{H}_s ,\varrho ]{\rm d}t + \gamma \mathcal{D}[\hat{a}] \varrho\: {\rm d}t
+ \sqrt{\gamma} \mathcal{H}[\hat{a}]\varrho \: {\rm d}w \:,
\ee
 where, for the sake of simplicity we have considered the case of zero temperature ($n_{\sf th}=0$), and we have 
 introduced the superoperator $\mathcal{H}[O]\varrho = O\varrho + \varrho O^\dag - \tr{\varrho (O+O^\dag)}\varrho$. \\
 
As anticipated in Sec. \ref{s:GaussianCPmap}, one can describe noisy Gaussian measurements, by applying a Gaussian dual map $\Phi^*$ (characterized by matrices $X^*$ and $Y^*$) to the measurement operators characterized by the covariance matrix $\sig_{m}$. {In the following we will focus on two noisy maps: loss evolution, that will lead to results equivalent to the ones obtained to describe inefficient photodetectors by using stochastic master equations and the unravelling matrix formalism \cite{WisemanDiosi}, and additive Gaussian noise, which is particularly easy to incorporate in our formalism.}
The matrices describing a lossy evolution are
\begin{align}
X = \sqrt{\eta} \mathbbm{1}_2 \:, \:\:\: Y =(1-\eta) \mathbbm{1}_2 \nonumber \:.
\end{align}
 Here $\eta \in [0,1]$ will denote the measurement efficiency, $\eta=1$ corresponding to no-loss and thus to a perfect detector. The matrices that characterise the dual map are thus
\begin{align}
 X^* &= X^{-1} = \mathbbm{1}_2 /\sqrt{\eta} \:, \nonumber \\
Y^* &= X^{-1} Y X^{-1 \sf T} = \frac{1-\eta}{\eta} \mathbbm{1}_2 \nonumber \:.
\end{align}
By evaluating the matrix
\be
\sig_{m}^* = X^* \sig_{m} X^{* {\sf T}} + Y^* \:,
\ee
one obtains the evolution Eqs.~(\ref{eq:diffusofiltering}) and (\ref{eq:diffusofilteringFM}), where $\sig_{m}$ is replaced with $\sig_{m}^*$. In particular, for the case considered above of continuous homodyne detection of the quadrature $\hat{x}$, one obtains the following matrices
\begin{align}
\tilde{A} = A + \left( 
\begin{array}{c c}
\frac{\eta\gamma (2 n_{\sf th}+1)}{\eta 2 n_{\sf th} +1} & 0 \\
0 & 0
\end{array}
\right) \:, \quad
\tilde{D} = D - \left( 
\begin{array}{c c}
 \frac{\eta\gamma (2 n_{\sf th} + 1)^2}{\eta 2 n_{\sf th} +1}  & 0 \\
0 & 0
\end{array}
\right) \:, \quad
B =
\left( 
\begin{array}{c c}
0 & \sqrt{\frac{\eta\gamma}{ \eta2 n_{\sf th} + 1}} \\
0 & 0
\end{array}
\right) \:. \label{lossmats}
\end{align}
{The analytical solution for the steady-state 
covariance matrix can be easily obtained in this case too. 
We report here the zero-temperature case ($n_{\sf th}=0$):
\begin{align}
\sig &= 
\left(
\begin{array}{c c}
\frac{\gamma(2\eta-1) - 2\chi + \sqrt{(\gamma+2\chi)^2 - 8\eta\gamma\chi}}{2\eta\gamma}& 0 \\
0 & \frac{\gamma}{\gamma - 2\chi}
\end{array}
\right) \:.
\end{align}
}
It is impressive
how such a comparatively simple formalism is capable of capturing quite a wide class of dynamics and monitoring processes.
The equivalent stochastic master equation for the density operator in this case would read
\be
{\rm d}{\varrho} =  - i [ \hat{H}_s ,\varrho ]{\rm d}t + \gamma \mathcal{D}[\hat{a}] \varrho\: {\rm d}t
+ \sqrt{\eta\gamma} \mathcal{H}[\hat{a}]\varrho \: {\rm d}w \:.
\ee

If we rather consider Gaussian additive noise, we have 
\begin{align}
X = X^* = \mathbbm{1}_2\:, \nonumber \\
Y = Y^* = \Delta \mathbbm{1}_2 \:.
\end{align}
and the conditional evolution of the covariance matrix is described by the matrices
\begin{align}
\tilde{A} = A + \left( 
\begin{array}{c c}
\frac{\gamma (2 n_{\sf th}+1)}{2 n_{\sf th} +1 + \Delta} & 0 \\
0 & 0
\end{array}
\right) \:, \quad 
\tilde{D} = D - \left( 
\begin{array}{c c}
 \frac{\gamma (2 n_{\sf th} + 1)^2}{2 n_{\sf th} +1 + \Delta}  & 0 \\
0 & 0
\end{array}
\right) \:, \quad
B =
\left( 
\begin{array}{c c}
0 & \sqrt{\frac{\gamma}{ 2 n_{\sf th} + 1 + \Delta}} \\
0 & 0
\end{array}
\right) \:. \label{darkmats}
\end{align}
{
Once again, the steady-state solution can be evaluated analytically, obtaining for $n_{\sf th}=0$ 
\begin{align}
\sig &= 
\left(
\begin{array}{c c}
\frac{\gamma(1-\Delta) - 2\chi(1+\Delta) + \sqrt{4\gamma^2\Delta + [(\gamma(\Delta-1) + 2\chi(1+\Delta)]^2}}
{2\eta\gamma}& 0 \\
0 & \frac{\gamma}{\gamma - 2\chi}
\end{array}
\right) \:.
\end{align}
This noisy detection corresponds to the stochastic master equation for the density operator $\varrho$ in the 
so called ``dark noise'' case \cite{WisemanMilburn}.} Note that, although the parameters $\Delta$ and $\eta$ (the detection efficiency from the previous case) represent distinct physical quantities, 
the class of conditional evolutions they describe are the same, as can be seen by setting $\Delta=(1/\eta-1)$ in Eq.~(\ref{darkmats}) and noticing that it gives rise to the same matrices $\tilde{A}$, $\tilde{D}$ and $B$ as in Eq.~(\ref{lossmats}).
It is also worth noticing that, if we consider a zero-efficiency detector, that is $\eta=0$ in the first example, or infinite noise added, that is the limit $\Delta\rightarrow\infty$ in the second example, one re-obtains the unconditional evolution of Eq. (\ref{diffusion}), with $\tilde{A}=A$, $\tilde{D}=D$ and $B=0$. The unconditional evolution does indeed correspond to continuous monitoring where all the measurement outcomes are discarded ($\eta=0$) or bring no information because of the infinite noise added ($\Delta\rightarrow \infty$). 
\subsection{\bf \textsf{Scattering induced diffusion }}
As one should expect, considering a different interaction between the system and the bath leads to a different open dynamics. 
In particular, let us set the following the interaction Hamiltonian 
\be
\hat{H}_C \:{\rm d}t = \sqrt{2\Gamma}\: \hat{x} \hat{x}_{in}^\prime \: dw \:.
\ee 
Note that, in quantum optics, 
it is customary to retrieve the interaction Hamiltonian considered in in Eq. (\ref{eq:interactionA}) of the previous example from the Hamiltonian above via rotating wave-approximation and under the assumption of weak-coupling. By also considering, for the sake of simplicity, the case of free Hamiltonian $\hat{H}=\omega (\hat{x}^2 + \hat{p}^2)/2$ for the system and a zero-temperature bath ({\em i.e.} $\chi=0$ and $n_{\sf th}=0$ from the previous example), one obtains the following drift and diffusion matrices governing the unconditional evolution:
\begin{align}
A = \left(
\begin{array}{c c}
0 & \omega \\
-\omega & 0 
\end{array}
\right)
 \:, \:\:\: D = \left(
\begin{array}{c c}
0 & 0 \\
0 & 2\Gamma 
\end{array}
\right),
\end{align}
that is, a {\em purely Hamiltonian} drift matrix (corresponding to no damping for the oscillator), and a diffusion matrix
where a momentum heating contribution is evident.
These equations can in fact be equivalently obtained from the following master equation
\be
\dot{\varrho} = -i [\hat{H},\varrho] + \Gamma \mathcal{D}[\hat{x}] \varrho \:, \label{eq:scatteringME}
\ee
which is known to describe, for example, the recoil heating of a dielectric nanosphere trapped in an optical cavity by optical tweezers \cite{Oriol}.
The unconditional dynamics of this system, characterised by the matrices $A$ and $D$ above, is not stable, 
in the sense of not admitting a steady state solution for the covariance matrix $\sig$.

However, including, as above, the continuous monitoring of the quadrature operator $\hat{x}$, one obtains (for perfect efficiency $\eta=1$ and zero additive noise, 
$\Delta=0$) a conditional evolution described by the matrices $\tilde{A}=A$, $\tilde{D}=D$ and 
\begin{align}
B = \left( \begin{array}{c c}
0 & \sqrt{2 \Gamma} \\
0 & 0
\end{array}
\right) \:.
\end{align}
Now, at variance with its unconditional diffusive counterpart, Eq. (\ref{eq:diffusofiltering}) with the $\tilde{A}$, $\tilde{D}$ 
and $B$ determined above does admit a steady state: 
it is thus shown that monitoring the environment allows one to stabilise the system. 
Besides, for measurements of unit efficiency, the steady state is always 
pure. This can be seen directly in our formalism -- by applying the formula for the purity (\ref{puri}) -- 
but it's also a general consequence of a fact which is manifest 
in the general quantum trajectory approach, that time-continuous projective measurements of the environment always keep the 
system state pure. Hence, in principle, the monitoring of the environment would allow one to purify ({\em i.e.}, essentially, to cool) the state of the system. 
Note that this is mere wishful thinking when an actual system, like a trapped levitating bead is 
considered, as the perfect monitoring of the environment would imply a perfect collection and detection of the light
scattered by the bead. 
It is still true, however, that even imperfect monitoring helps substantially in the endeavour of cooling and squeezing a 
trapped levitated bead \cite{bead}.

Once more, let us emphasise that the same findings could be obtained by starting from the stochastic master equation
\be
{\rm d} {\varrho} = -i [\hat{H},\varrho] \: {\rm dt} + \Gamma \mathcal{D}[\hat{x}] \varrho \: {\rm d}t + \sqrt{\Gamma}\mathcal{H}[\hat{x}] \varrho \: {\rm d}w \, ,
\ee
that describes the conditional evolution of a mechanical oscillators whose position is continuously monitored 
via the scattered light \cite{doherty99}.

\section{\bf \textsf{Summary and Conclusions}} \label{s:conclusions}

In this article, we have introduced the notion of Gaussian quantum states and the associated formalism, discussed in detail 
the characterisation of Gaussian CP-maps and Gaussian (general-dyne) measurements, and then moved on to 
consider dynamics resulting from linear interaction with a white noise environment. Concerning the latter, we covered both 
the unconditional dynamics, resulting from discarding the degrees of freedom of the environment, and the conditional ones arising 
when the environment is continuously measured through general-dyne detection. Such conditional dynamics ask for the introduction 
of quantum Wiener processes and we hence referred collectively to all these situations as ``diffusive'' dynamics, in analogy 
with classical stochastic mechanics. 

A number of interesting side results cropped up along the way: the complete characterisation of Gaussian CP-maps, 
while already known \cite{demoen},
was never proven as compactly as in Appendix \ref{appa}; the brief discussion of dual Gaussian CP-maps 
is also original; the possibility of deriving full master equations by only considering 
the open Gaussian dynamics was never explicitly pointed out, again to the best of our knowledge. 

Above all, our derivation of the conditional Riccati equation (\ref{briccone}) only makes use of the 
update of the quantum state as prescribed by standard quantum mechanics. 
We hope that the physics community may find such a derivation 
easier to access than treatments based on other forms of filtering, which are more common in the 
literature on quantum stochastic processes and calculus \cite{james}.
Also, while equivalent forms were well known and have been used extensively over the last thirty years of research 
in quantum mechanics, quantum optics and quantum stochastic processes, we would also like to remark that 
the parametrisation through the measurement covariance matrix $\sig_m$ provided in Eq.~(\ref{briccone}) is novel. 
We find it to be particularly compact and expedient when compared to alternative options, and it is our belief that 
its immediate connection to a well defined detection scheme might prove of benefit to the portion of the 
physics community with an interest in Gaussian processes.

\vspace*{.4cm}

\subsubsection*{\bf \textsf{Acknowledgments}}

A.~Doherty and H.~Wiseman contributed, through very insightful discussions over the past two years, 
to our understanding of several key points related to the material covered in this article.
The form (\ref{briccone}) of the Riccati equation was derived during the 
Workshop on Quantum Control Engineering 
held at the Newton Institute, University of Cambridge in July and August 2014. 
The mathematical argument behind the characterisation of Gaussian CP-maps given in Appendix \ref{appa}
emerged while AS was lecturing LL at Scuola Normale Superiore (Pisa), in the spring of 2014.
MG and AS acknowledge financial support from EPSRC through grant EP/K026267/1. LL acknowledges the support from the Spanish MINECO Project No. FIS2013-40627-P and the CIRIT Project No. 2014 SGR 966 of the Generalitat de Catalunya.\bigskip
\appendix

\section{\bf \textsf{Complete characterisation of deterministic Gaussian CP-maps}}\label{appa}
We report here the proof that any pair of real matrices $X$ and $Y$ satisfying (\ref{heisdyn}) 
correspond to a deterministic Gaussian CP-map. Throughout the appendix, we shall liberally 
refer to formulae from the main text and Section \ref{s:GaussianCPmap}.

As will be apparent {\em \`a posteriori}, restricting to the case $m=2n$ (number of modes of the environment equal to 
twice the number of modes of the system) and $\sig_E=\id$ (zero temperature environment) will suffice to 
reproduce all of the dynamics in question. Such assumptions will be hence made in the following. 

Eq.~(\ref{XYAB}) shows that, given $X$ and $Y$, $A$ may reproduce the Gaussian CP-map through 
a symplectic reduction if and only if $A=X$. The choice of $B$ allows instead for some freedom: by setting $\sig_E=\id$ 
and fixing the dimension of $B$ as discussed above one may write 
\be
B = \sqrt{Y} O \; , \label{BO}
\ee
with $O$ is a $2n\times 4n$ real matrix with orthornormal rows. 
Note that, since $i\Omega -iX\Omega X^{\sf T}$ is anti-symmetric and hence yields no contribution 
if contracted with a real vector, Inequality (\ref{heisdyn}) implies $Y\ge 0$, ensuring the existence of $\sqrt{Y}$.
We must then find an $O$ such that $A$ and $B$ satisfy the symplecticity condition (\ref{symplect}), {\em i.e.} such that 
\be \label{XOY}
X\Omega_n X^{\sf T} + \sqrt{Y} O\, \Omega_{2n} O^{\sf T} \sqrt{Y} = \Omega_n \; .
\ee

It is now extremely useful to restrict to the case of a strictly positive $Y$. 
This can be done by noticing that, if one shows an $O$ can be found for each $Y>0$ such that (\ref{XOY}) is fulfilled, then, 
for any $Y\ge 0$, one has that for all $\varepsilon > 0$ there exists an $O_\varepsilon$ such that
\be
X\Omega_n X^{\sf T} +\sqrt{Y+\varepsilon \id}\, O_\varepsilon\, \Omega_{2n}\, O_\varepsilon^{\sf T} \sqrt{Y+\varepsilon \id}\, = \Omega_n 
\ee
(true since $Y+\varepsilon \id>0$). As the set of matrices $O$'s with orthogonal rows is compact, since it can inherit the topology of the 
compact set $O(4n)$, in the limit $\varepsilon\rightarrow 0$ there exists a converging subsequence of $O_\varepsilon$ whose limit $O_0$ is also contained in the set of possible $O$'s.
Then one can apply the $\varepsilon\rightarrow 0$ limit on that subsequence to the equation above and obtain 
\be
X\Omega_n X^{\sf T} +\sqrt{Y}\, O_0\, \Omega_{2n}\, O_0^{\sf T} \sqrt{Y}\, = \Omega_n \; .
\ee  
We can hence restrict the remainder of the proof to the case of a positive definite $Y$.

The matrix $Y^{1/2}$ may then be assumed to be invertible, and the condition (\ref{XOY}) may be recast as
\be
i O \Omega O^{\sf T}  =  iY^{-1/2} \left(\Omega - X\Omega X^{\sf T}\right) Y^{-1/2}  \ge -\id \, ,
\ee 
where we incorporated the CP-map condition (\ref{heisdyn}) in the last inequality. 

The anti-symmetric $Y^{-1/2} \left(\Omega - X\Omega X^{\sf T}\right) Y^{-1/2}$ may be brought, through the action by similarity 
of an orthogonal $R\in O(2n)$, to the canonical form:
\be
R Y^{-1/2} \left(\Omega - X\Omega X^{\sf T}\right) Y^{-1/2} R^{\sf T} = \left( \begin{array}{cc}
0 & D \\
-D & 0 
\end{array}\right) \, , \label{anticano}
\ee
with a diagonal $D={\rm diag}(d_1\,\ldots,d_n)$ satisfying $0 \le D\le \id$.\footnote{\label{anticanote}The canonical decomposition of anti-symmetric matrices follows from the diagonalisability of symmetric ones: let $A$ be a real, anti-symmetric, $2n\times2n$ matrix (even dimension is just imposed to fix ideas and because it applies to our case), then $A^2$ is symmetric and can be diagonalised as per $O A^2 O^{\sf T} = B$, with $B$ diagonal and $O\in O(2n)$. Consider then a generic eigenvector ${\bf e}_1$ of $A^2$, with eigenvalue $d_1^2\in {\mathbbm R}$. The vector ${\bf e'}_1=A {\bf e}_1/d_1$ is clearly orthogonal to ${\bf e}_1$, because $A$ is antisymmetric: ${\bf e}_1^{\sf T} A{\bf e}_1=0$. Let ${\bf v}$ be a generic vector in the linear subspace orthogonal to the space spanned by ${\bf e}_1$ and ${\bf e'}_1$, then one has
$$
{\bf v}^{\sf T}A {\bf e_1} = {\bf v}^{\sf T} {\bf e_1'}d_1  = 0  \quad {\rm and} \quad {\bf v}^{\sf T}A {\bf e_1'} = {\bf v}^{\sf T}A^2 {\bf e_1}/d_1 = 0 \; , 
$$
as ${\bf e_1}$ is an eigenvalue of $A^2$ by hypothesis. 
The equation above shows that any choice of orthogonal basis including ${\bf e}_1$ and ${\bf e'}_1$ would result in $A$ acting as a diagonal block $d_1\omega$ in the subspace spanned by ${\bf e}_1$ and ${\bf e'}_1$. Iterating this argument leads to the canonical decomposition  
applied in Eq.~(\ref{anticano}).}
The latter inequality ($d_j\le 1$ $\forall j$) is inferred from the condition (\ref{heisdyn}). 
The matrix $Y^{-1/2} \left(\Omega - X\Omega X^{\sf T}\right) Y^{-1/2}$ can be decomposed in the direct sum of separate 
anti-symmetric two-dimensional blocks, each of which can be obtained from the $4\times 4$ matrix $\Omega_2$ through 
the $2\times4$ matrix $O$ chosen as follows:
\be
O = \left(\begin{array}{cccc}
\cos\theta_j & 0 & 0 & -\sin\theta_j \\
0 & \cos\theta_j & \sin\theta_j & 0 
\end{array}
\right) \, ,
\ee
so that, by setting $\sin(2\theta_j)=d_j$, one has
\be
O O^{\sf T} = \left(\begin{array}{cc}1 & 0 \\ 0& 1 \end{array}\right) \quad {\rm and }\quad 
O \left(\begin{array}{cc}0& \id_2\\-\id_2&0\end{array}\right) O^{\sf T} = \left(\begin{array}{cc}0 & d_j \\ -d_j& 0 \end{array}\right)\, .
\ee

We have thus proven that, for all real $X$ and $Y$ satisfying (\ref{heisdyn}), there exist real $A$ and $B$ such that 
$X=A$, $Y=BB^{\sf T}$ and $A\Omega A^{\sf T}+B\Omega B^{\sf T}=\Omega$.
The latter matrix equation can be recast by stating that the $2n$ vectors ${\bf v}_j$ forming the rows of the matrix $(A\; B)$ verify 
${\bf v}_j\, \omega\, {\bf v}_k^{\sf T} = \Omega_{jk}$, where $\omega=\left( \begin{smallmatrix} \Omega_n & 0 \\ 0 & \Omega_{2n} \end{smallmatrix}\right)$ is the symplectic product on the bipartite system.
It is left to show that matrices 
$C$ and $D$ exist such that $S = \left(\begin{array}{cc}A & B \\ C & D\end{array}\right)$ is symplectic, 
which is equivalent to stating that one can extend such vectors ${\bf v}_j$ to a global symplectic basis for $\omega$.

This can always be accomplished, as for any $2N\times 2N$ anti-symmetric matrix $\omega$, any set of $2s$ row vectors ${\bf v}_j$
such that ${\bf v}_j\omega{\bf v}^{\sf T}_{k}=\left(\Omega_s\right)_{jk}$ for $j,k\in[1,\ldots,2s]$, 
where $\Omega_s$ is the $2s\times2s$ standard symplectic form of $s$ modes, 
can be completed to a basis of $2N$ linearly independent vectors such that 
${\bf v}_j\omega{\bf v}^{\sf T}_{k}=(\Omega_N)_{jk}$ for $j,k\in[1,\ldots,2N]$.\footnote{This is the case since 
the first $2s$ vectors may be completed to a linearly independent basis and then 
orthogonalised with respect to the symplectic product according to the 
mapping ${\bf v}_j \mapsto {\bf v}_j - \sum_{k,l \le 2s} (\Omega_s)_{k l} {\bf v}_{j}\omega{\bf v}_{l}^{\sf T}\, {\bf v}_{k}$, for all $j\in [2s+1,\ldots,2N]$. 
One then just needs to rotate the added orthogonal vectors to bring the antisymmetric matrix of their symplectic products in the canonical 
form, see footnote \ref{anticanote}, and then rescale them to achieve a symplectic basis. See Theorem 1.15 of \cite{Gosson} for details.}

This completes the proof that each pair of $X$ and $Y$ that verify (\ref{heisdyn}) correspond to a deterministic Gaussian CP-map,
resulting from the reduction of a symplectic dynamics on a larger space.

Besides, we have also shown that such a reduction may always be constructed by considering an environment of $2n$ degrees of freedom 
(where $n$ are the degrees of freedom of the system) initially in a pure state (as the CM $\sig_E=\id$ has determinant $1$ and is hence
associated with a pure Gaussian state).

\section{\bf \textsf{The dual of a Gaussian CP-map\label{app:dual}}}

We derive here the action of a dual Gaussian CP-map as given in text in Eqs.~(\ref{dual1}-\ref{dual2}), and also provide the reader with some additional remarks and mathematical details about dual Gaussian CP-maps. In what follows, we denote by $\Phi^{*}$ the dual of a Gaussian CP-map $\Phi$ whose action is determined through Eqs.~(\ref{eq:CPmap1st}-\ref{eq:GaussianCPmap}) by two matrices $X,Y$, of which $X$ is assumed to be invertible. The Stinespring representation of $\Phi$ allows us to write for an arbitrary operator $A$
\begin{equation*}
\Phi(A)\ =\ {\rm Tr}_2\, [\,\hat{U} \, A\otimes \sigma_0\, \hat{U}^\dag ]\ ,
\end{equation*}
where $\sigma_0$ is the vacuum state of an environment and $\hat{U}$ is a symplectic unitary verifying
\begin{align*}
\hat{U}^\dag \hat{{\bf R}}\, \hat{U}\ &=\ S\, \hat{{\bf R}}\; , \\
\hat{U} D_{\bf R} \hat{U}^\dag\ &=\ \hat{D}_{S {\bf R}}\; ,
\end{align*}
${\bf R}$ being a real vector appearing in a displacement operator of the bipartite system,
defined as per Eq.~(\ref{eq:weyl}). 
Moreover, we know from appendix~\ref{appa} that $S$ is represented in block form as
\begin{equation*}
S\ =\ \begin{pmatrix} X & \sqrt{Y}\,O \\ * & * \end{pmatrix} \; ,
\end{equation*} 
with $*$ standing for generic, undefined matrix blocks.
Now, by the very definition of dual map, we have
\begin{equation*}
{\rm Tr}_1\,[A\, \Phi^{*}(B)]\ =\ {\rm Tr}_1\,[\Phi(A) B]\ =\ {\rm Tr}_{1,2}\, [\, \hat{U}\, A\otimes \sigma_0\; \hat{U}^\dag\, B\otimes\mathds{1}\, ]\ =\ {\rm Tr}_1\, [A \ {\rm Tr}_2[\mathds{1}\otimes \sigma_0\ \hat{U}^\dag B\otimes\mathds{1}\, \hat{U} ]\, ]\ .
\end{equation*}
Since $A$ is generic, this means that for arbitrary $B$ we have
\begin{equation}
\Phi^{*} (B)\ =\ {\rm Tr}_2\, [\,\mathds{1}\otimes \sigma_0\ \hat{U}^\dag B\otimes\mathds{1}\, \hat{U}\, ]\; .
\end{equation}
Applying this formula with $B=\hat{D}_{\Omega {\bf r}}$ gives us
\begin{equation}
\begin{split}
\Phi^{*}(\hat{D}_{\Omega {\bf r}})\ &=\ {\rm Tr}_2\, [\,\mathds{1}\otimes \sigma_0\ \hat{U}^\dag \hat{D}_{\Omega {\bf r}}\otimes\mathds{1}\, \hat{U}\, ]\ =\ {\rm Tr}_2\, [\,\mathds{1}\otimes \sigma_0\ \hat{U}^\dag \hat{D}_{\Omega ({\bf r}\oplus {\bf 0})}\, \hat{U}\, ]\ =\\ 
&=\ {\rm Tr}_2\, [\,\mathds{1}\otimes \sigma_0\ \hat{D}_{S^{-1} \Omega ({\bf r}\oplus {\bf 0})}\, ]\ =\
{\rm Tr}_2\, [\,\mathds{1}\otimes \sigma_0\ \hat{D}_{\Omega S^{\sf T} ({\bf r}\oplus {\bf 0})}\, ]\ =\\ 
&={\rm Tr}_2\, [\,\mathds{1}\otimes \sigma_0\ \hat{D}_{(\Omega X^{\sf T} {\bf r}\, \oplus\, \Omega O^{\sf T} \sqrt{Y} {\bf r})}\, ]\ =\ \hat{D}_{\Omega X^{\sf T} {\bf r}} \ {\rm Tr}\, [\,\sigma_0\, \hat{D}_{\Omega O^{\sf T} \sqrt{Y} {\bf r}}\, ]\ =\\
&=\ \hat{D}_{\Omega X^{\sf T} {\bf r}} \ \chi_0(-\,\Omega O^{\sf T} \sqrt{Y} {\bf r})\ =\ \hat{D}_{\Omega X^{\sf T} {\bf r}} \ e^{-\frac{1}{4} {\bf r}^{\sf T} Y {\bf r}}\; .
\end{split}
\end{equation}
The action of the dual CP-map $\Phi^{*}$ on a generic Gaussian state $\varrho_G$ (assumed to have vanishing first moments) can also be easily determined, 
for invertible $X$, by applying the equation
above on the Fourier-Weyl expansion of $\varrho_G$, given by inserting a Gaussian characteristic function into Eq.~(\ref{multifw}).
\begin{align}
\Phi^{*}\left(\varrho_G\right) &= \frac{1}{(2\pi)^n}\int_{{\mathbbm R}^{2n}} \hspace*{-.3cm}{\rm e}^{-\frac14{\bf r}^{\sf T}\sig{\bf r}}\: 
\Phi^{*}\left(\hat{D}_{\Omega {\bf r}}\right) {\rm d}^{2n}{\bf r} \nonumber \\
&= 
\frac{1}{(2\pi)^n}\int_{{\mathbbm R}^{2n}} \hspace*{-.3cm}{\rm e}^{-\frac14{\bf r}^{\sf T}(\sig+Y){\bf r}} 
\hat{D}_{\Omega X^{\sf T}{\bf r}}\, {\rm d}^{2n}{\bf r} \nonumber \\
& = 
\frac{1}{(2\pi)^n|{\rm Det} X|}\int_{{\mathbbm R}^{2n}} \hspace*{-.3cm}{\rm e}^{-\frac14{\bf r}^{\sf T}(X^{-1} \sig X^{-1 \sf T}+X^{-1}Y X^{-1 \sf T}){\bf r}} \hat{D}_{\Omega {\bf r}}\, {\rm d}^{2n}{\bf r} \, .
\end{align}

Note that ${\rm Tr}\left[\Phi^{*}\left(\varrho_G\right)\right]=\frac{1}{|{\rm Det}X|}$. 
The channel $\Phi^{*}$ is hence a Gaussian  completely positive map, but not a trace preserving one unless $|{\rm Det}X|=1$. 
Since a quantum channel is trace preserving if and only if its dual is unital,
it follows that a trace-preserving Gaussian CP-map $\Phi$
with $X$ invertible is 
unital if and only if $|{\rm Det}X|=1$.

\end{document}